# Survey of $CO_2$ radiation experimental data in relation with planetary entry

Philippe Reynier[1]

*ISA, 33610 Cestas, France*

This paper focuses on a survey of experimental data related to radiation into $CO_2$ plasma flows, which are encountered during Mars and Venus entries. The review emphasizes on VUV and IR radiation, since recent experimental efforts has been devoted to these wavelength ranges. The main objective of the study is to identify the most attractive datasets for future crosscheck comparisons with the results obtained during future test campaigns with ESTHER shock-tube. The survey accounts for the results obtained in shock-tubes, expansion tube and plasma arc-jets for Mars and Venus test campaigns. The experimental results obtained for propulsion related studies have also been considered.

## 1 Nomenclature

**Acronyms:**

ADST:   Arc-Driven Shock Tube

CEV:    Crew Exploration Vehicle

DLAS:  Diode Laser Absorption Spectroscopy

EAST:  Electric Arc Shock Tube

HVST: High Velocity Shock Tube

IR:        InfraRed

ISTC:   International Science and Technology Center

JAXA:  Japan Aerospace eXploration Agency

LIF:     Laser Induced Fluorescence

---

[1] Research Engineer, ISA (Ingénierie et Systèmes Avancés), 33610 Cestas, France, Philippe.Reynier@isa-space.eu

MIPT: Moscow Institute of Physics and Technology

MSL: Mars Surface Laboratory

NASA: National Aeronautics and Space Administration

NIR: Near InfraRed

TALIF: Two-photon Absorption Laser Induced Fluorescence

TPS: Thermal Protection System

TRP: Technology and Research Programme

VIS: VISible

VUV: Vacuum UltraViolet

## 2 Introduction

Radiation measurements in $CO_2$ plasma flows are of interest for different aerospace applications. In propulsion, their investigation is necessary to study the combustion process in engines and to analyse exhaust plumes. Concerning atmospheric entries, they are key issues for planets such as Mars and Venus, since $CO_2$ is a major component of their atmospheres. Radiation from species formed by $CO_2$ dissociated flows, covers a wide range of wavelengths since some systems radiate in the vacuum ultraviolet (VUV) and other in the infrared.

Measurements in VUV and infrared bands cover a range of wavelength of 62 up to 200 nm, and beyond 700 nm respectively. In hypersonic flows, radiative heating can be significant in both of these spectrum bands depending on plasma composition. However, VUV emission measurement is difficult to achieve due to the instrumentation limitations. Usually dedicated deuterium lamps [1] are used for such measurements. These devices required a specific calibration based on advanced photon metrology [2]. VUV measurements are not only relevant for hypersonic applications, but are also of interest for nuclear fusion, photochemistry, and exobiology; more particularly infrared measurements are a key issue for propulsion systems.

In the following, a literature survey of available data for $CO_2$ radiating flows is presented. First, the data related to propulsion applications are presented, then experimental data obtained for Venus and Mars entry conditions are surveyed. The survey activity benefits from previous studies carried out for Mars entry [3] and shock facilities [4], as well as shock-tube data for Mars entry [5]. The available heritage from previous ESA TRP activities [6-7] has also been considered. In the frame of a research effort on CFD validation in $CO_2$ environment, a large spectral database for $CO_2$ radiation has been gathered, evaluated, and incorporated in the radiation tool PARADE [8-9]. The survey is mostly focusing on the results obtained in shock-tubes and expansion tubes. Data from plasma torches have been also considered; available spectra and measurements could be of interest even if they have been measured for much different flow conditions than those encountered within a shock-tube. The objective of the current study is to identify the most attractive datasets for $CO_2$ radiation, in line with planetary entry, with a focus on VUV and IR contributions. These datasets will be of interest for assessing the future experimental data to be obtained during test campaigns in the European shock tube ESTHER [10]. ESTHER is a facility developed by an international consortium led by IST of Lisbon, under funding from the European Space Agency for supporting future exploration missions. It is a two-stage combustion driven shock-tube (see Figure 1) with laser ignition, with the capability to reach a shock velocity up to 14 km/s for Earth atmospheric entry. Specific instrumentation is going to be added to the facility, with a VUV capability for high-speed entries and a specific IR instrumentation for testing Mars entry conditions. The goal of the current effort is to support the development of this IR capability.

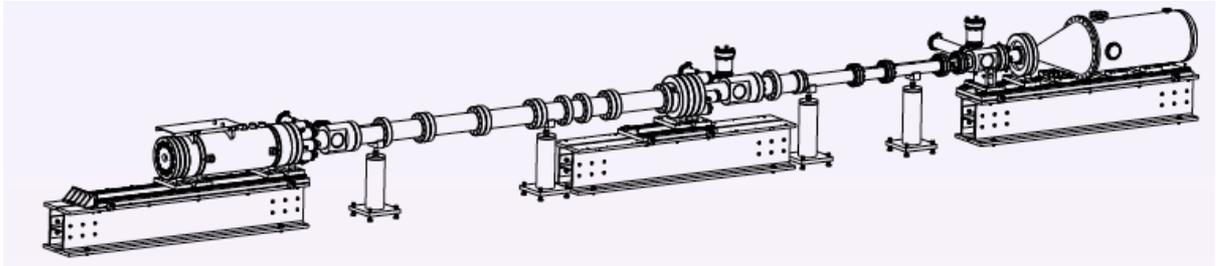

Figure 1: Sketch of ESTHER shock-tube (credit IST)

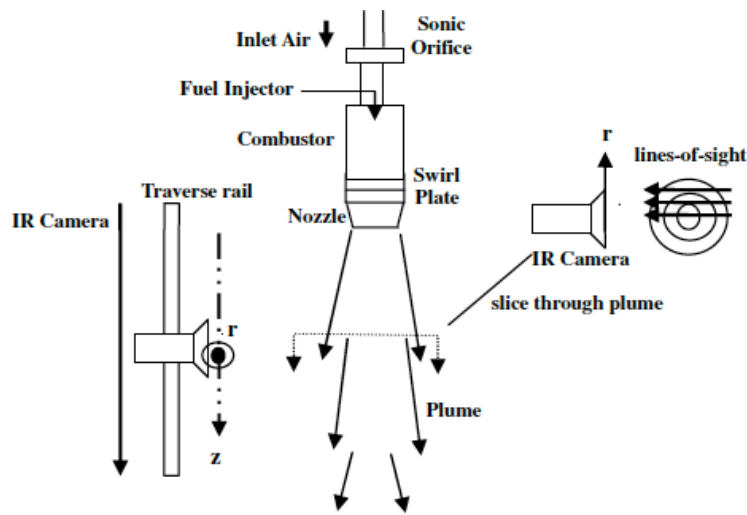

Figure 2: Set-up for investigating radiation in exhaust plumes (from [14])

## 3 Propulsion analysis

### 3.1 Introduction

The exhaust plume of an aircraft, or a rocket, is one of the sources of the object IR signature. As a consequence, radiation plumes identification and tracking are of interest for missile and aircraft systems. Species such as $CO_2$, $CO$, $H_2O$, as well as other combustion products [11] have to be searched for estimating radiation, particularly when species related systems and bands are in the IR domain.

Another interest of $CO_2$ radiation for propulsion applications is $CO_2$ expansion flows [12] in nozzles. This is relevant for sample return mission to planets with a $CO_2$ atmosphere such as Mars or Venus (more particularly to Mars since the ground pressure is low contrary to Venus for which a launch from a balloon could be envisaged). For this objective, activities related to modelling [13], database comparison [11], and experiments [12,14-15] can be found in the literature. In the following, the emphasis will be given to recent experimental campaigns carried out for $CO_2$ radiation.

## 3.2 Experiments

A generic arrangement for investigating plumes is shown in Figure 2. One of the purposes of this test campaign [14] was to investigate the narrowband radiation intensity emitted from a plume. Exit temperature of the exhaust gases was in the vicinity of 800 K. Radiation intensity measurements were obtained using an infrared camera. Corresponding two-dimensional mapping of the radiation intensity distribution at different moments and time-averaged are shown in Figure 3. The two-dimensional mapping radiation measurements obtained this study is of high interest for future instrumentation developments in hypersonic shock-facilities. In this study, narrowband radiation calculations were performed to estimate the contribution from heated water vapour and carbon dioxide to the measured intensity.

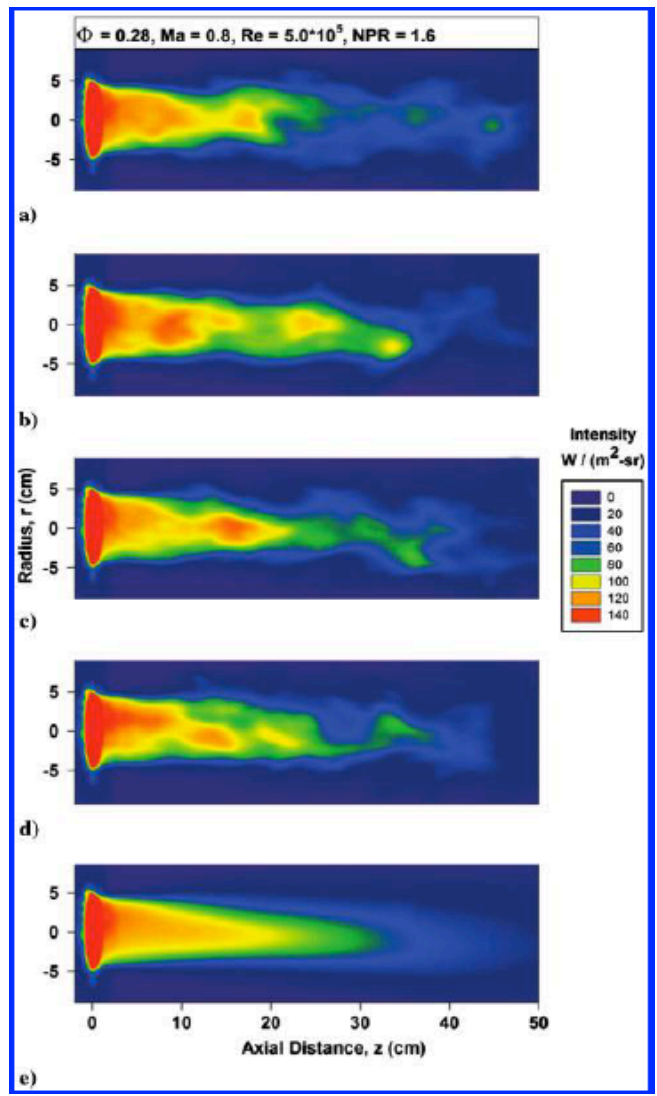

**Figure 3: Infrared radiation intensity measurements in a plume (Mach 0.28, exit temperature 800 K): (a)-d) instantaneous (elapsed time of 12 ms); e) averaged. (from [14])**

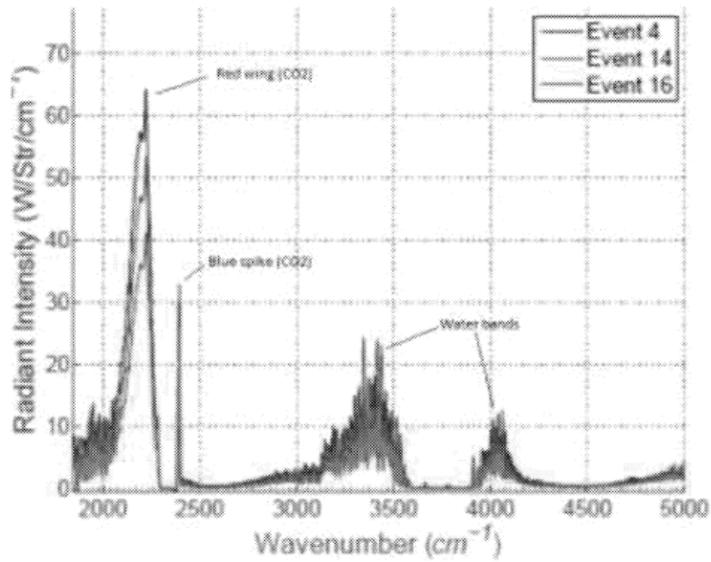

**Figure 4: Plume spectral intensity in the 2000-5000 cm$^{-1}$ range (from [15])**

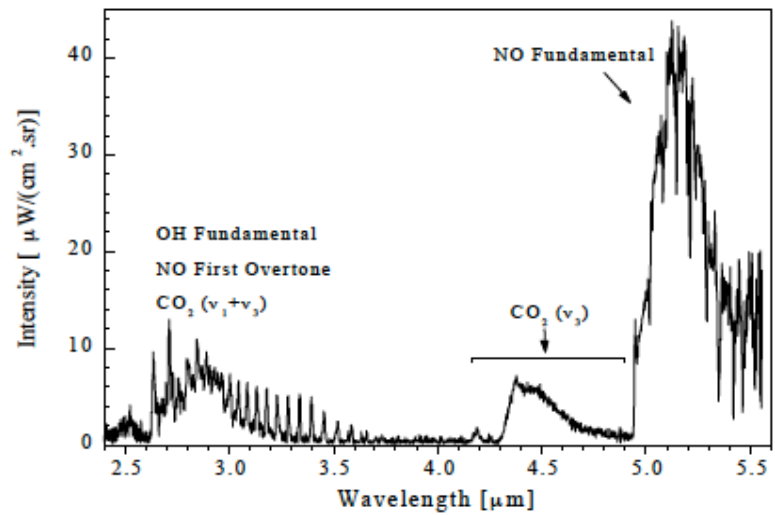

**Figure 5: Measured IR spectrum at 1 atm and 3400 K (from [16])**

A similar experimental study on plume infrared signature was recently conducted by Higgins et al [15], but for at higher temperatures, above 2000 K. These authors performed infrared radiation emission measurements of a liquid engine plume, measuring plume radiant intensity profiles, plume IR image, and plume spectral emission distribution. An example of the results obtained, with water and $CO_2$ measured bands is shown in Figure 4.

In the same context of infrared signature measurements, Packan et al [16] have carried out an extensive experimental campaign on radiation in low temperature air plasma. For the tests, air was containing small quantities of $CO_2$ and $H_2O$, measurements were performed in a 50 kW inductive plasma torch at 3400 K. Measured emitted spectrum is shown in Figure 5, highlighting the contributions of NO, OH, CO, and $CO_2$ bands.

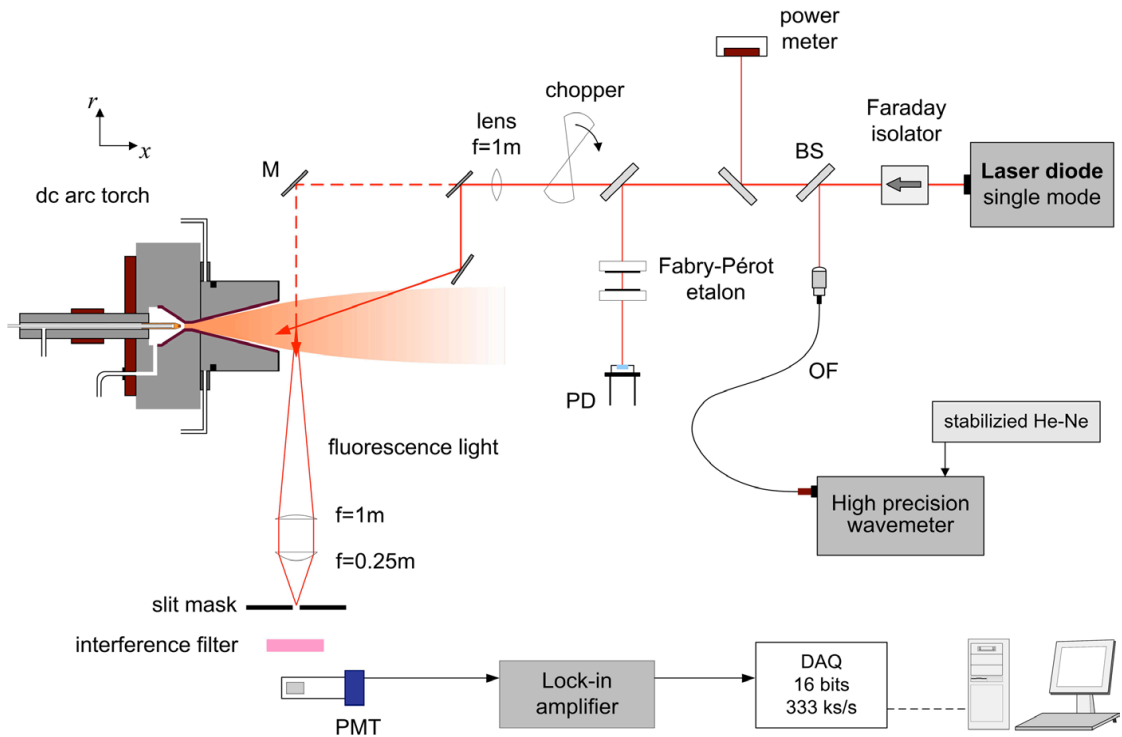

**Figure 6: Experimental arrangement for $CO_2$ expansion flows (from [12])**

| Pressure (Torr) | Velocity (km/s) | Reference |
|---|---|---|
| 0.5 | 11.4 | 1, 17, 15 |
| 0.5 | 10.6 | 1, 17, 15 |
| 1 | 9.5 | 1, 17, 15 |

**Table 1: Venus entry conditions investigated for VUV data measurements (1 Torr = 133 Pa)**

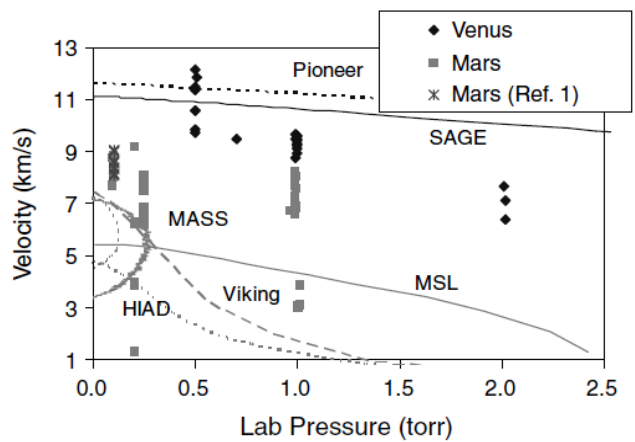

**Figure 7: Test matrix for Mars and Venus retained in [17] (1 Torr = 133 Pa)**

In the perspective of sample return missions to Mars, nozzle flows of $CO_2$ have to be extensively investigated, since the launch of the return capsule from Mars ground will be performed in a $CO_2$ atmosphere. The same would apply for Venus, but for a much higher pressure, even if a launch at high altitude using a balloon could be envisaged. For such objective, $CO_2$ expansion flows have been studied by Mazouffre & Pawelec [12] in an arc-jet operating at low pressure, and both pure $CO_2$ and $CO_2$-$N_2$ mixture were investigated for flow enthalpy of 9.4 and 6 kJ/kg respectively. The experimental set-up is schematized in Figure 6. Tests were performed for supersonic plasma jet, at 4200 m/s for $CO_2$, and 3400 m/s for $CO_2$-$N_2$, with temperature of 17 000 K and 11 000 K in the shock region respectively. Measurements were based on the excitation of oxygen atom metastable level $O(^5S)$ as the laser frequency is scanned over the oxygen triplet system (3s → 3p transition) at 777.1944 nm; however they were mostly focused on the velocity and temperature distributions within the plasma jet and no radiation spectra have been published.

## 4  Venus

### 4.1  Test conditions

Several experimental studies have been focused on radiation measurements for Mars aerocapture and Venus entry conditions. Experimental campaigns for Venus entry have been carried out in EAST shock-tube by Martinez [1], and Cruden et al [17-18]. The calibration of the deuterium lamp selected for the VUV measurements is detailed in [1]. In these studies, a 96.5 $CO_2$ and 3.5 $N_2$ atmosphere was considered for Venus, while a 96 $CO_2$ and 4 $N_2$ atmosphere was retained for investigating a Mars aerocapture in [1]. The different entry conditions investigated for VUV measurements related to Venus entry are resumed in Table 1. A comparison between the conditions tested in [17] and the some entry probe trajectories is shown in Figure 7. Experiments were performed from the VUV to IR ranges, from 120 up to 1700 nm at high velocities, in [1], and up to 5000 nm for low velocities, in [17-18].

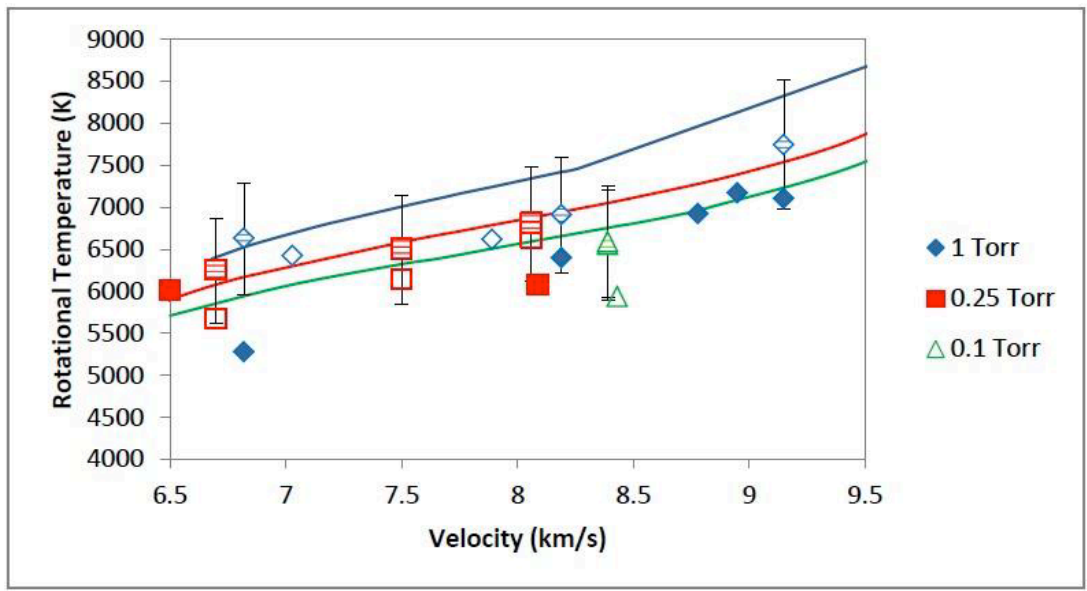

Figure 8: Rotational temperature for CN and C2 systems (from [1])

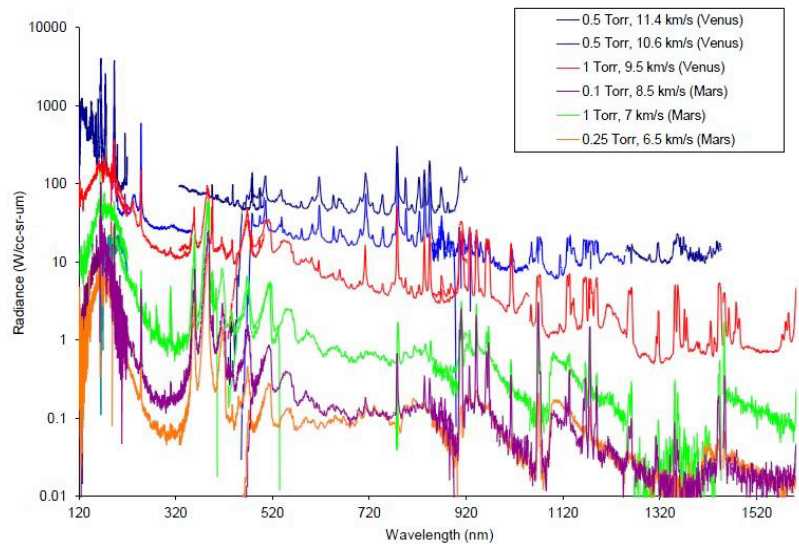

Figure 9: Mars and Venus spectra obtained in [1]

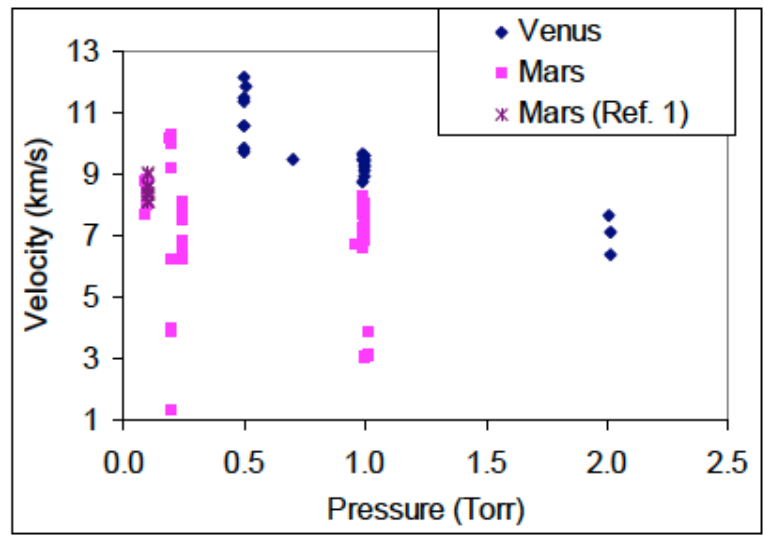

Figure 10: EAST test conditions for Venus and Mars entries [17]

**4.2 Available datasets**

The study carried out by Martinez [1] focuses more particularly on CN, and $C_2$ and their related rotational temperatures. The results obtained for the rotational temperature as function of the shock velocity are displayed in Figure 8. They highlight the increase of the rotational temperature with the shock velocity. The spectra from 120 to 1520 nm, obtained by Martinez [1] for both Mars aerocapture, at 6.5, 7, and 8.5 km/s, and Venus entry conditions, at 9.5, 10.6, and 11.4 km/s, are shown in Figure 9. Datasets cover VUV up to NIR radiation, with the major contribution in the VUV. The VUV range is dominated by the CO(4+) system, two carbon atomic lines are also present at 193 and 248 nm.

Other test campaigns for Venus entry conditions (3-12 km/s 0.1-2 Torr) were conducted in EAST [17-18] spectral radiance was investigated from VUV through mid-IR (120-1650 nm, and 3000-5000 nm). Tests conditions are plotted in Figure 10. Resolved spectra of the CO $4^{th}$ positive band in the VUV was also reported for the first time. Measurement of $CO_2$ molecular vibrational radiation was also attempted for low velocities. The contributions of the different spectral regions from VUV to IR, to the total radiance for Venus and Mars entry conditions are resumed in Figure 11. Mid IR was only investigated at low velocities (3-4 km/s). For the other measurements medium resolution imaging was performed for $C_2$ Swan, CN Violet, O 777 nm triplet and C 193 and 248 nm lines.

The observed results are similar to those reported in [1] that can be seen in Figure 9. One of the main radiator is CN radiating 22% less in Venus mixture than in Mars mixture, all other parameters being equal.

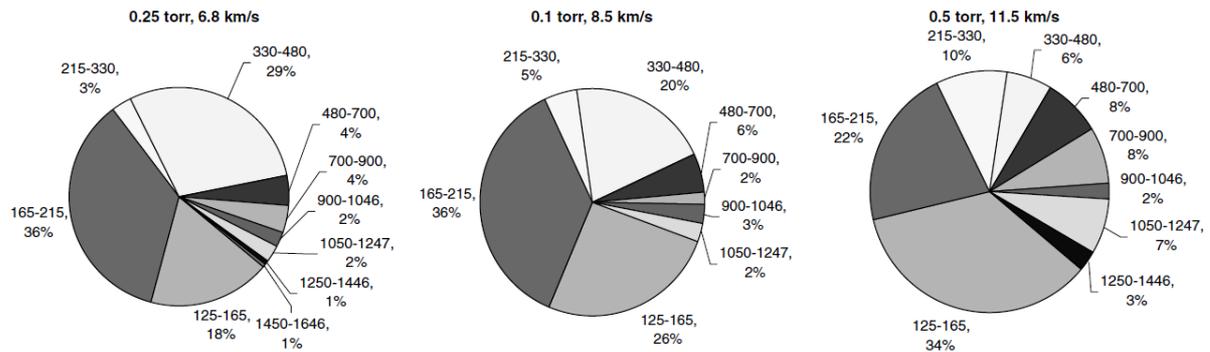

**Figure 11: Contributions to the different spectral regions to the total radiance for different Mars (96% $CO_2$, 4% $N_2$) and Venus (96.5% $CO_2$, 3.5% $N_2$) entry conditions [17]**

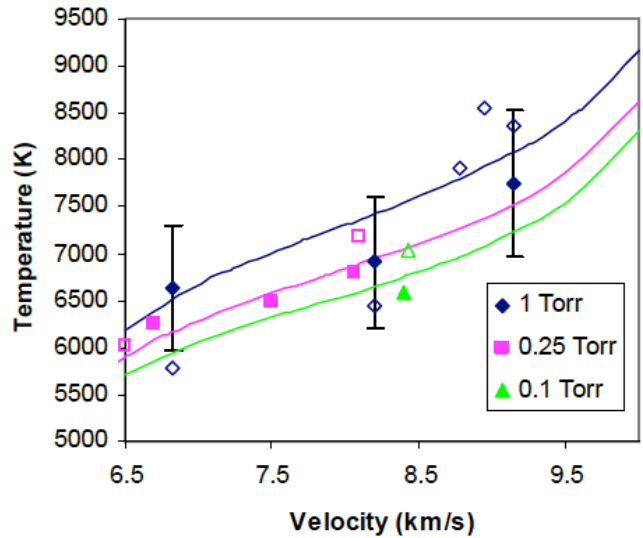

**Figure 12: Rotational temperature for Venus and Mars conditions [17]**

The analysis of the different tests shows that below 10 km/s, there is a strong emission from 120-300 nm, due to the CO $4^{th}$ positive system reaching a maximum at 160 nm, this is the dominant feature in the spectrum. Other major contributors to radiation emission are, NO, strong atomic lines of C, O, and N, and CN Violet, with the four bands at 359, 388, 422, and 461 nm, corresponding to CN Violet system. The radiation intensity emitted by the CN Violet system, increases with the velocity up to 10 km/s, then decreases and vanishes, since the energy present in the flow is sufficient to dissociate the molecule. Swan band for $C_2$, in between 500 and 700 nm, is also present in Figure 9, but at a much weaker level than CN. Like for CN, $C_2$ contribution disappears above 10 km/s due to dissociation effects. The same phenomenon applies to the CN red system, which is strong in the NIR above 700 nm below 9 km/s and vanishes when the shock velocity increases. With the increase in velocity, the atomic lines become visible, as highlighted in Figure 9, with the oxygen atomic triplets at 777 and 1130 nm, and the C atomic lines and 193, 248, and 1069 nm. However, the main features when increasing the velocity is the growth of the background continuum radiation certainly due to interactions between the electrons and atomic and molecular species that include free-bound and bremsstrahlung processes.

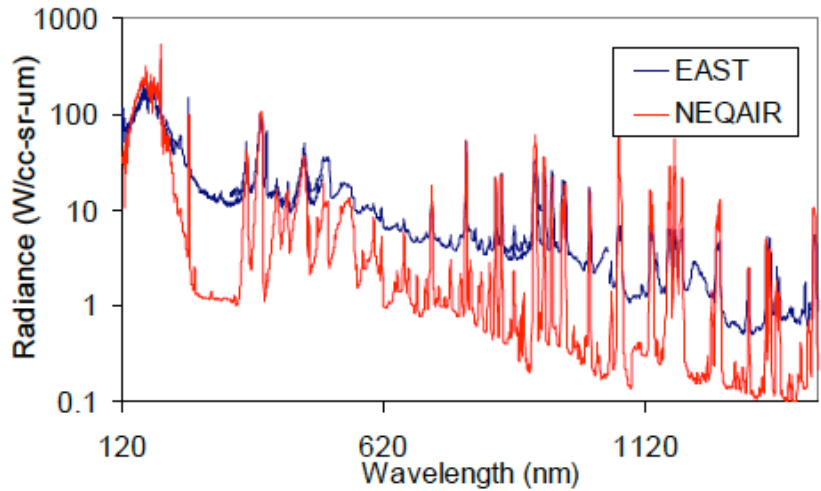

**Figure 13: Comparison between EAST data and NEQAIR for Venus entry at 1 Torr and 9.5 km/s [17]**

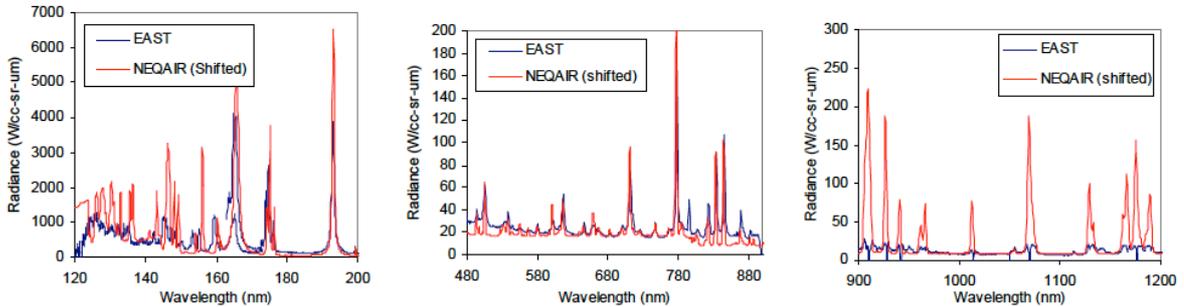

**Figure 14: Comparison between EAST data and NEQAIR for Venus at 0.5 Torr: a) VUV spectrum at 11.4 km/s; b) VIS-NIR spectrum at 10.5 km/s; c) NIR spectrum at 11.4 km/s [17]**

The contributions to the total radiance for different wavelength bands have been computed by Cruden et al [17] and are shown in Figure 11. The first contribution for all pressures and velocities is from VUV to far UV wavelengths, in between 125-215 nm, with 45 to 58% of the total radiation. At these wavelengths the radiation is mostly produced by the CO $4^{th}$ positive system and from NO. The IR radiation above 900 nm is small with around 10 % of the total radiation. The radiation in the visible domain can be mainly attributed to the CN red system. The region in between 330 and 480 nm, contributes to 35% of the total radiation at 6.5 km/s. This strong contribution decreases at higher velocities, with 12 % at 11.5 km/s due to the dissociation effects that are strong. Radiation from $C_2$ Swan, 480-700 nm is for the different cases lower than 10 % of the total radiation.

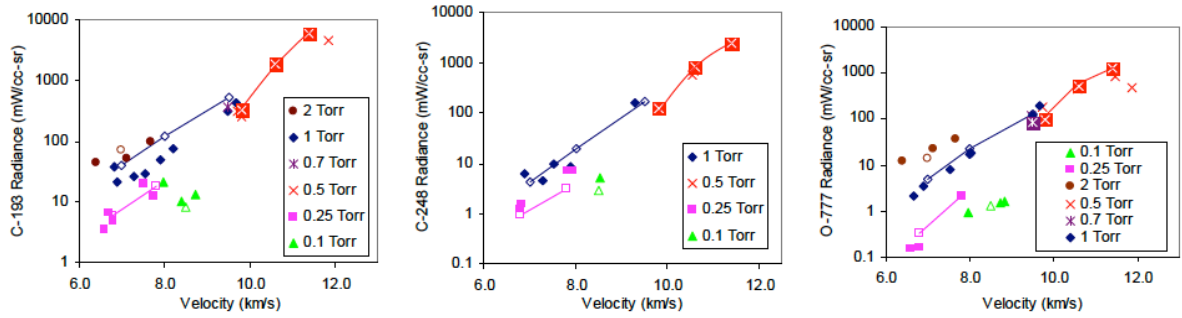

**Figure 15: Integrated radiance from carbon lines (193 and 248 nm) and oxygen triplet (777 nm) (EAST data as solid points; NEQAIR results as lines) [17]**

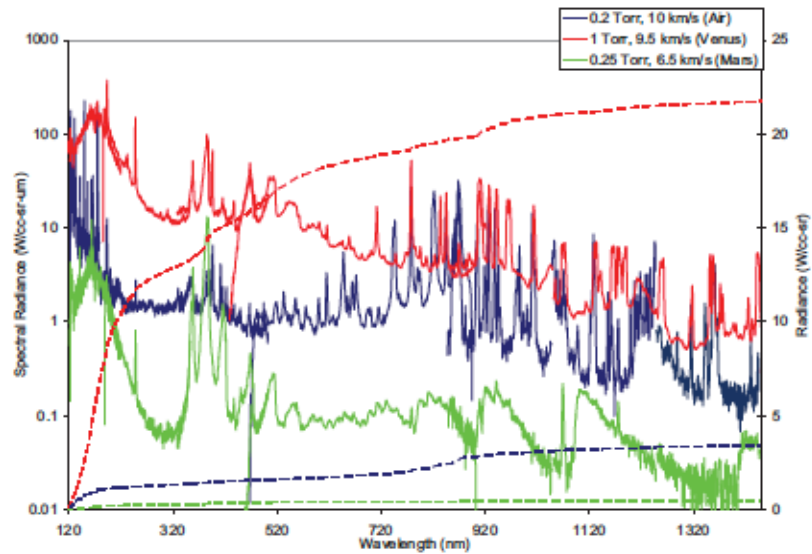

**Figure 16: Equilibrium radiation spectra and total radiative flux (dashed lines) for Earth, Venus and Mars entry conditions [20]**

Measurements of $C_2$ and CN bands systems have been used to compute rotational temperature; corresponding results are plotted in Figure 12. Finally, the tests have been reconstructed with NEQAIR [19]. Comparisons between experimental data and numerical calculations for different spectra are shown in Figures 13 and 14. Figure 15 shows the comparisons between EAST data and NEQAIR calculations, of the integrated radiance from carbon lines and oxygen triplet, for the different studied conditions.

Radiation emission spectra from VUV to mid-IR for Earth, Venus, and Mars entry conditions have been compared in [20] they are plotted in Figure 16. In the same contribution, Cruden [20] has provided extensive details on the calibration of the different measurement techniques used for the test campaigns including the deuterium and mercury lamps for VUV and UV-VIS measurements respectively. Mercury lamp is used for spectral calibration using atomic line sources in UV-VIS, in the infrared typically Hg lamps are used for the same finality. For calibration below 300 nm, a deuterium source is well adapted since the other lamps suffer form ambient absorption.

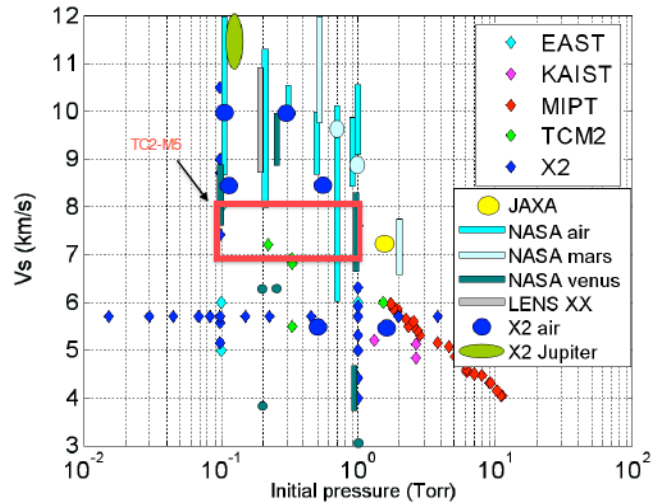

**Figure 17: TC2 test conditions and facilities maps [22]**

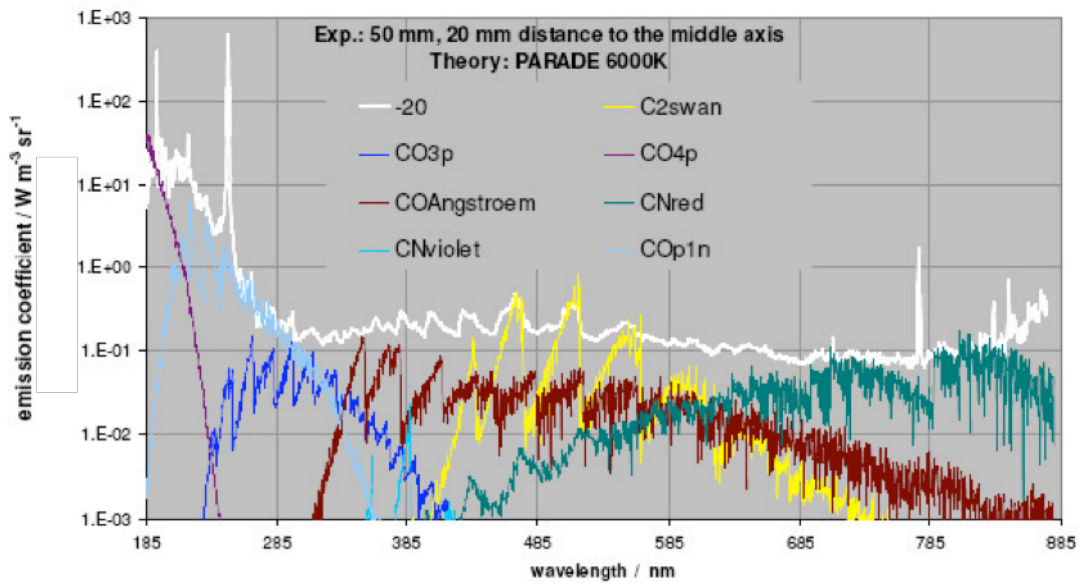

**Figure 18: Emission spectra (in white) in PWK-3 at 21 MJ/kg and PARADE calculations with contributions from different systems [25]**

## 5 Mars entry

### 5.1 ESA activities heritage

Since the Mars Premier project developed by CNES [3], ESA and CNES have fostered the activities of the European Radiation Working Group for improving radiation capabilities within Europe. In the frame of the working group activities, a test case has been dedicated to Mars entry [21]. The test case conditions are a pressure in between 0.1 and 1 Torr, a velocity range from 7 to 8 km/s and a nominal Mars atmosphere with 96% of $CO_2$ and 4% of $N_2$. They are reported in Figure 18 where the maps of different facilities as well as some existing datasets have been included [22].

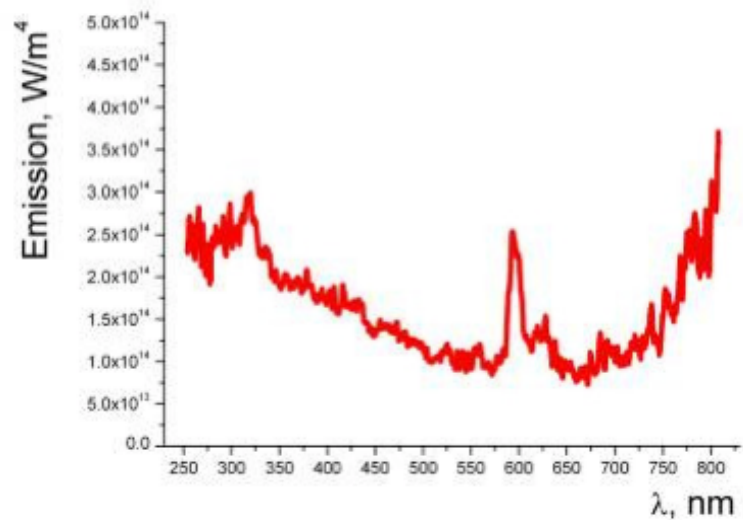

**Figure 19: Tests performed in VUT-1: Emission spectrum obtained for 4050 m/s and 11.1 Torr [25]**

In the frame of its TRP Programme, ESA had funded the development of the PARADE [23] code dedicated to radiation calculations, and of two research activities focused on the CFD validation in $CO_2$ environment [24-25]. Several test campaigns were performed during the first activity [24] in arc-jets, plasmatron, and hot shot facilities using DLAS, LIF, TALIF, and other conventional techniques but no data on radiation and chemical kinetics of interest were mentioned in the final report of this study, even if emission spectroscopy measurements were performed.

In a parallel study [25], tests have been conducted in arc-jets and in the VUT-1 MIPT shock-tube. A test campaign was performed in IRS PWK-3 plasma inductive wind-tunnel for Mars flow enthalpies of 10, 15, and 20 MJ/kg. Emission spectroscopy measurements were carried out during these tests for UV and visible ranges, from 175 to 880 nm. Atomic lines for carbon and oxygen were clearly observed as well as molecular emission from $C_2$ Swan, $CO^+$(1-) and CO (3+), as highlighted in Figure 18, in which the numerical reconstruction with PARADE [23] is also reported. It has to be noted that the numerical reconstruction was performed before the upgrade of the $CO_2$ radiation database done in [9]. In the frame of the same study, shock-tube tests were performed in VUT-1 [25] and pure $CO_2$ for a velocity range from 4 to 5.7 km/s, and for pressures from 100 to 1000 Pa. Both absorption and emission measurements were carried out from 230 to 810 nm has shown in Figure 19. During the same test campaign electron density was also measured using a microwave diagnostic technique, the results obtained are summarized in Figure 20.

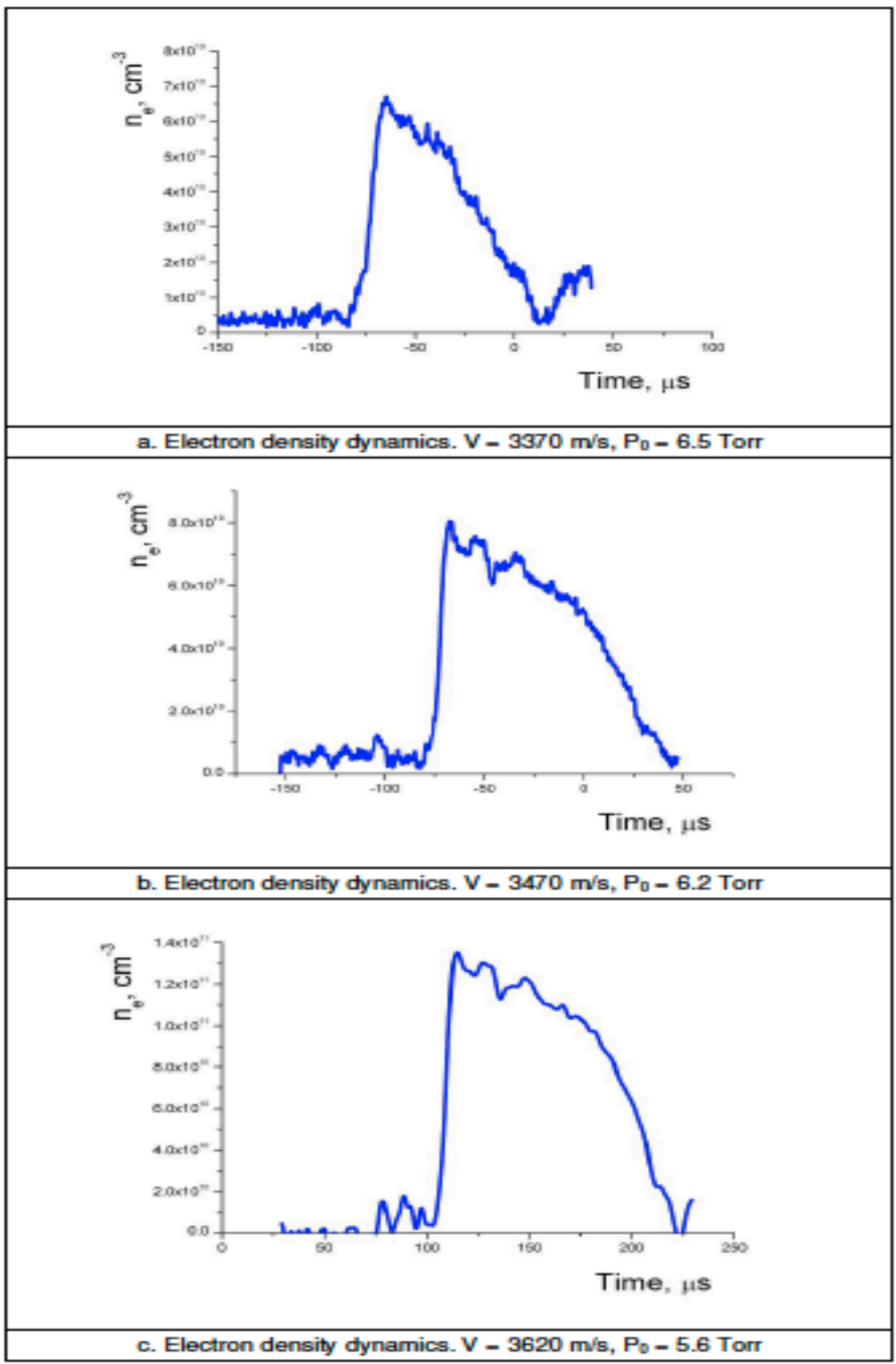

**Figure 20: Measured (using a streak camera) electron number densities in VUT-1 [25]**

## 5.2 EAST

Among the available facilities such as shock and expansion tubes, for investigating radiation and chemical kinetics for Mars entry, EAST has been the most used and extensive test campaigns carried out for a wide range Mars entry conditions. Some of these test conditions have been summarized by Cruden [26] in Figure 21.

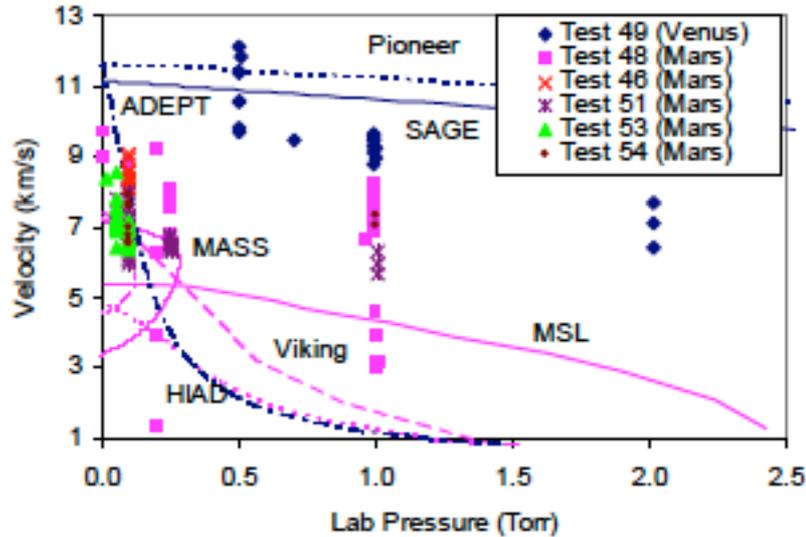

**Figure 21: Summary of tests carried out in EAST for Mars entry from 2008 to 2012 (from [26])**

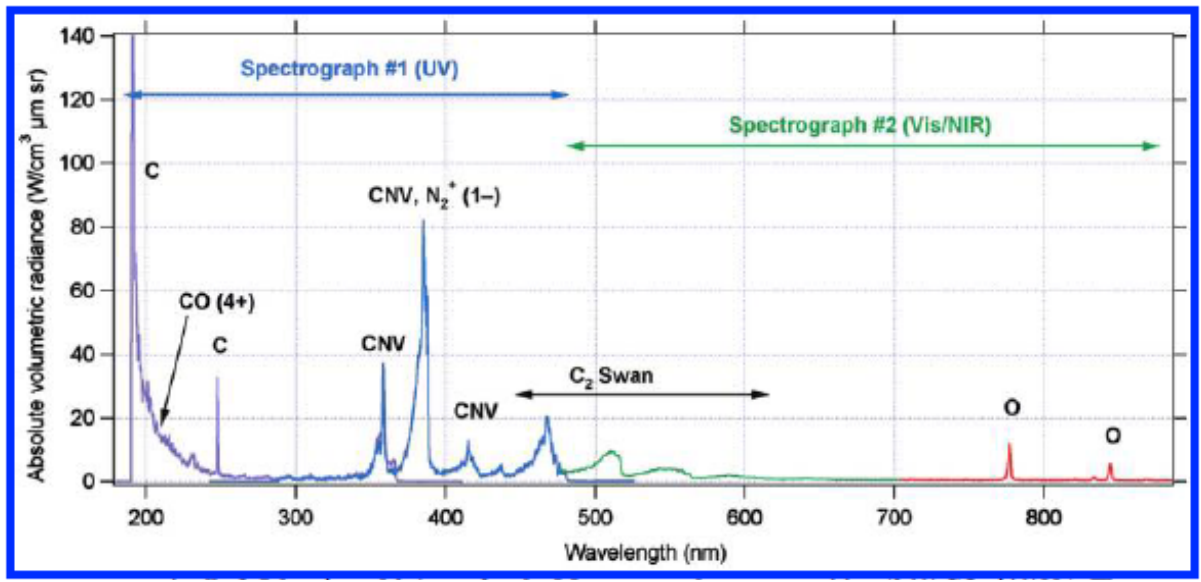

**Figure 22: Emission spectrum for a Mars atmosphere (96% $CO_2$, 4% $N_2$) at 13.3 Pa and a shock velocity of 8.6 km/s (from [27])**

The high velocity range, corresponding to Mars aerocapture conditions, has been studied by Grinstead et al [27] who have investigated the radiation and species kinetic relaxation for a Mars atmosphere at 0.1 Torr and a shock velocity of 8.6 km/s. Emission spectrum from VUV up to NIR has been measured and is displayed in Figure 22. The major source of radiance is from CO(4+) system even if CN violet, $N_2^+$ (1-) and $C_2$ Swan systems as well as several atomic lines for carbon and oxygen are present. High-resolution spectra with corresponding spatial variation of radiance for CN violet and $C_2$ Swan bands have been also published.

Detailed VUV part of the spectrum is shown in Figure 23.

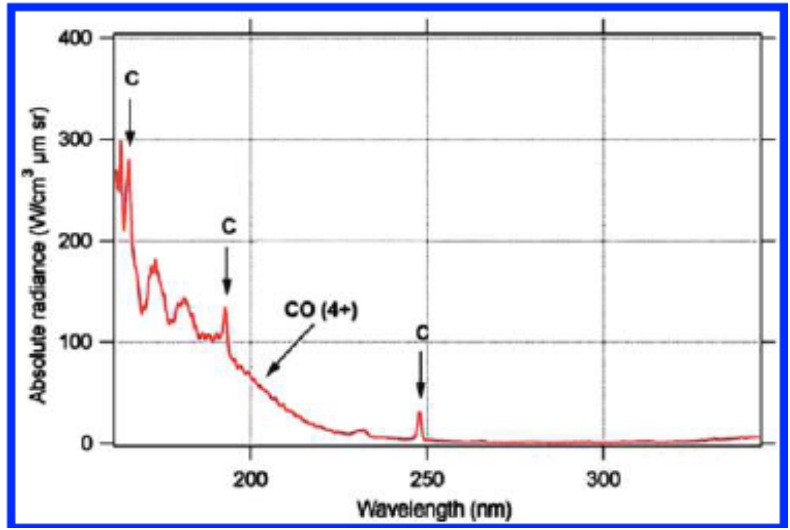

**Figure 23: VUV part of the radiance spectrum for Mars at 0.1 Torr (13.3 Pa) and 8.6 km/s (from [27])**

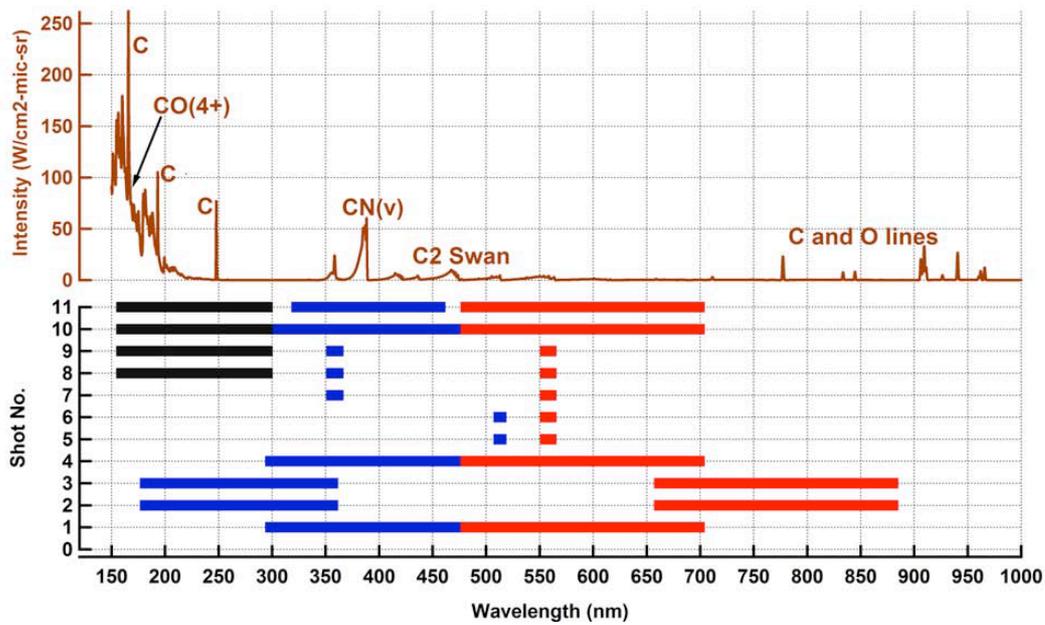

**Figure 24: Predicted spectrum and wavelength coverage for the different shots using three spectrographs (from [28])**

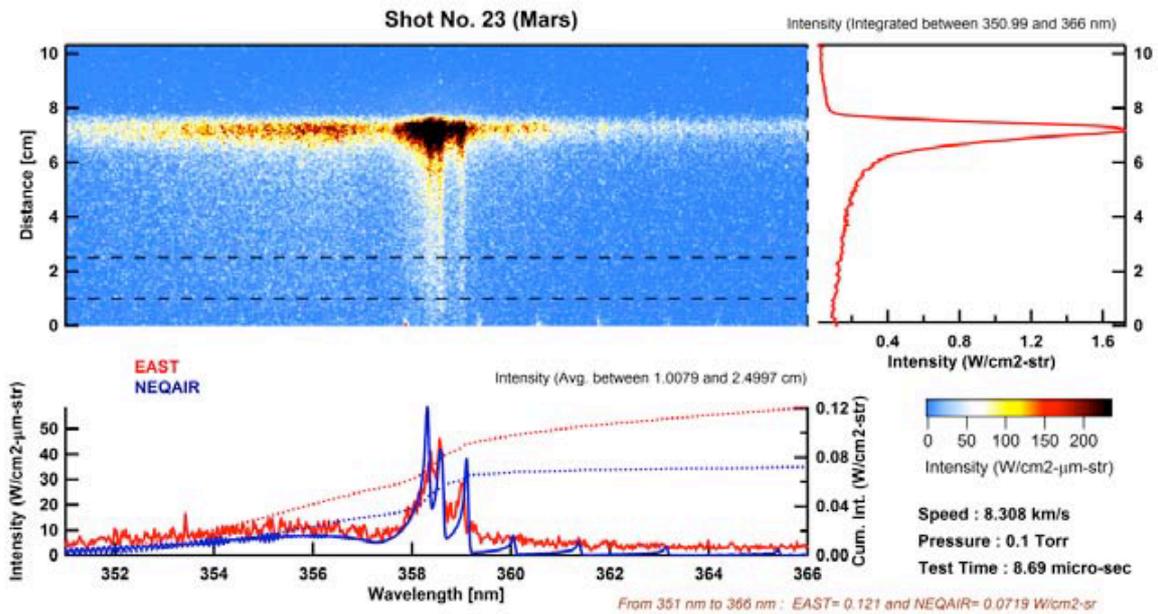

**Figure 25: Comparison between NEQAIR and EAST data in the UV range, at shock velocity of 8.55 km/s and a pressure of 0.1 Torr. The dotted curves are the cumulated intensity: measured in red and computed in blue (from [28])**

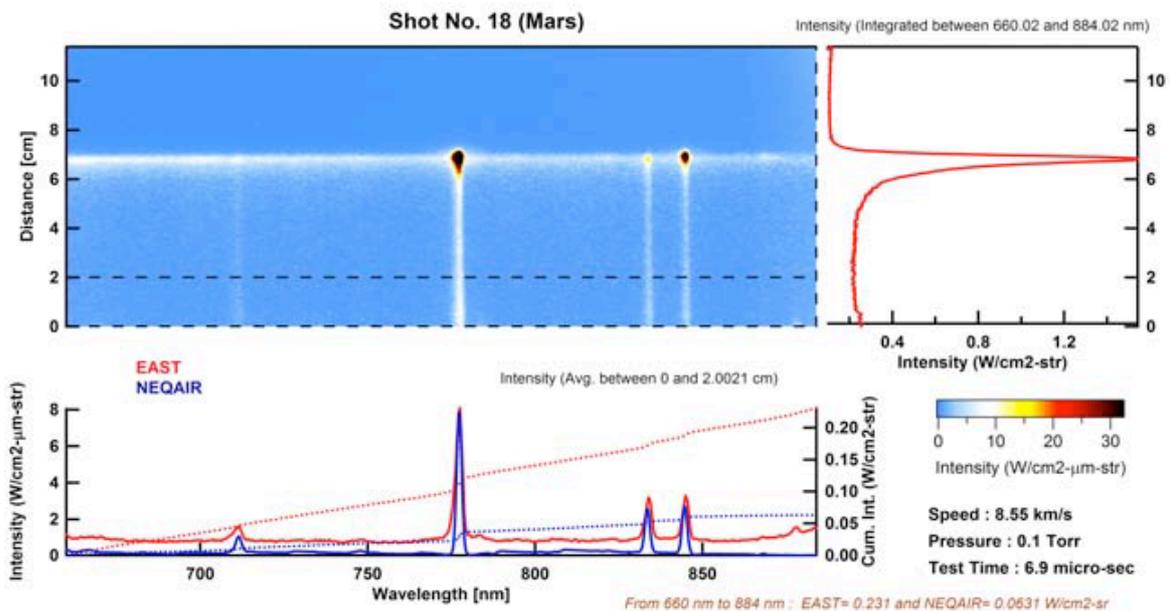

**Figure 26: Comparison between NEQAIR and EAST data in the NIR range at 0.1 Torr and a shock velocity of 8.55 km/s. The dotted curves are the cumulated intensity: measured in red and computed in blue (from [28])**

In another contribution, Bose et al [28] present some experimental data obtained for similar conditions, same pressure and temperature, a Mars atmosphere composition, and shock velocities ranging from 8.3 to 9 km/s. Three spectrographs were used for the test campaign covering wavelengths from the VUV to the NIR as shown in Figure 24. Test results have been compared with NEQAIR predictions; some of these comparisons are plotted in Figures 24 to 27 for shock velocities of 8.3, 8.55, 8.63, and 9 km/s respectively.

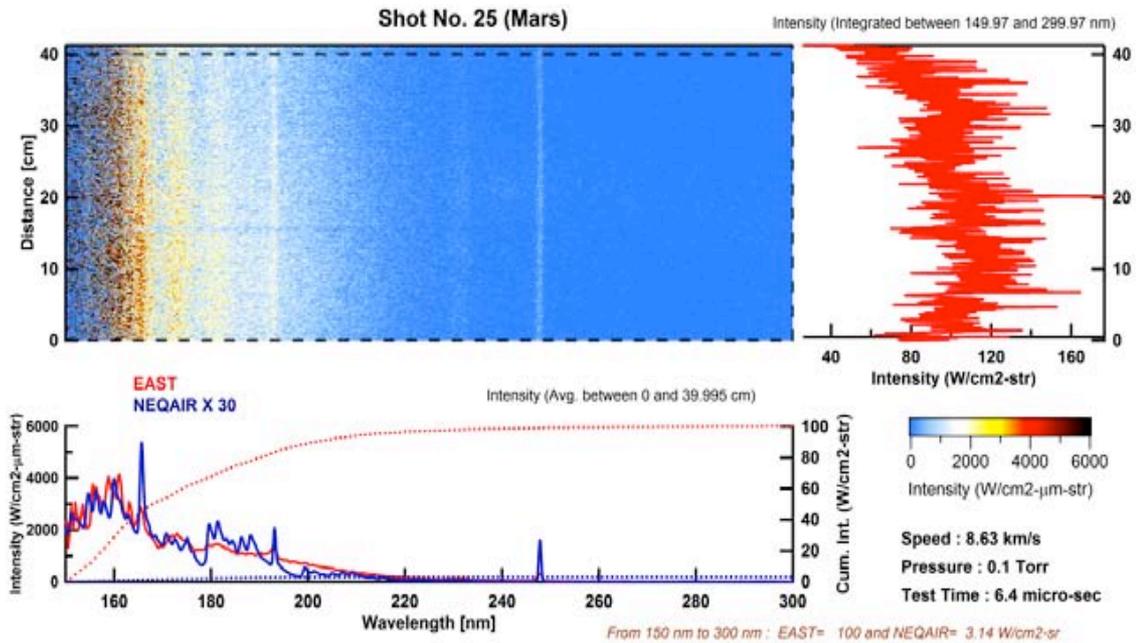

**Figure 27: Comparison between NEQAIR and EAST data in the VUV range at 0.1 Torr and a shock velocity of 8.63 km/s. The dotted curves are the cumulated intensity: measured in red and computed in blue (from [28])**

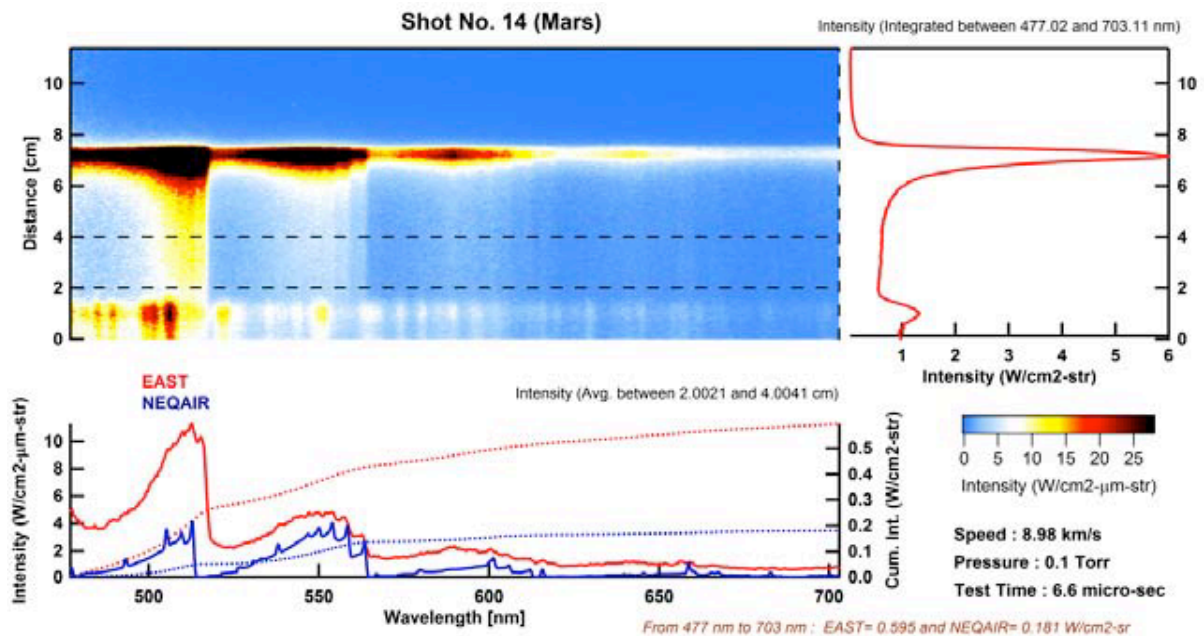

**Figure 28: Comparison between NEQAIR and EAST data in the VIS at 0.1 Torr and a shock velocity of 9 km/s. The dotted curves are the cumulated radiative intensity: measured in red and computed in blue (from [28])**

The Test Case 2 of the European Radiation Working Group has been also run in EAST [26]. Test Case nominal conditions are a nominal Mars atmosphere with 96 % of $CO_2$ and 4 % of $N_2$, velocities of 7, 7.5, and 8 km/s, and pressures of 0.1 and 1 Torr. Different runs were performed in EAST in these ranges and are reported in Figure 29 with the wavelength ranges covered by the instrumentation. The colour correspond to the data quality: yellow = low, dark green = medium, bright green = high. Measurements were performed from VUV to mid IR.

Composite spectra for the nominal conditions of Test Case 2 are plotted in Figure 30 in a log scale. The emitted radiation increases by a factor in between 2 and 3, from 7 to 8 km/s, has highlighted in the same figure. As a consequence, if for most of the probe entries performed so far, radiation was not the key issue for the TPS design, this will not be the case if the aerocapture of a probe has to be performed.

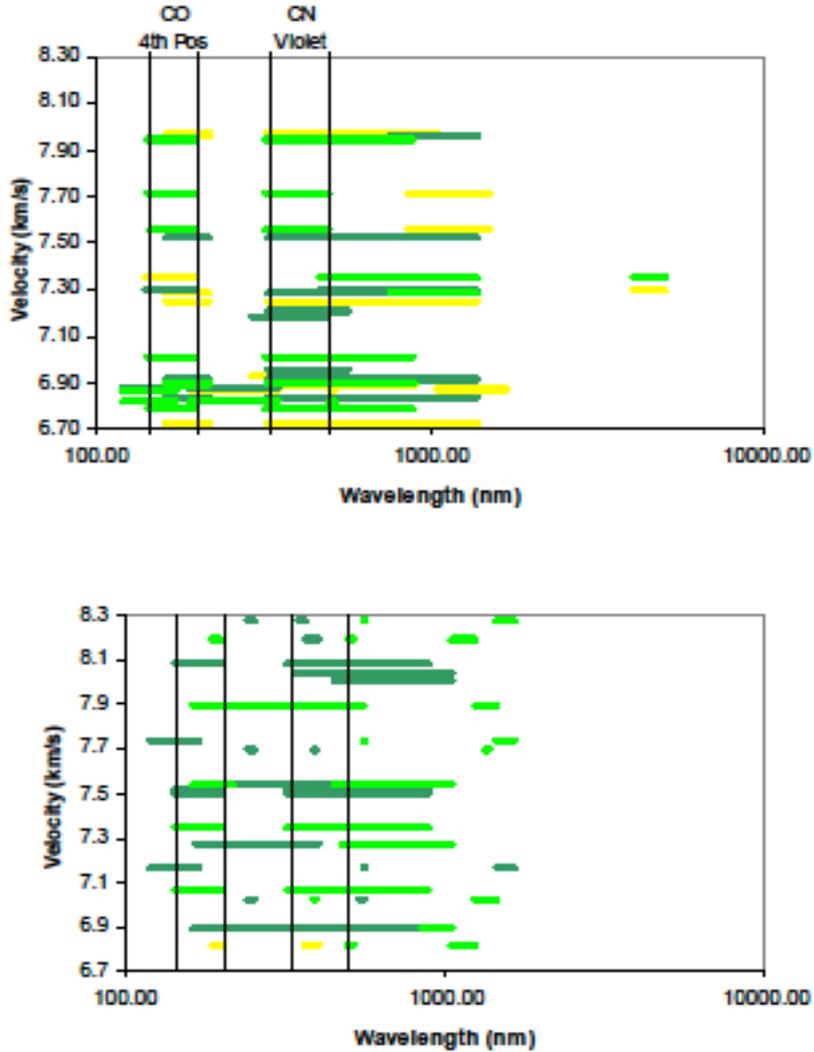

**Figure 29: Spectral ranges covered during EAST tests (bottom 1 Torr, top 0.1 Torr (from [26])**

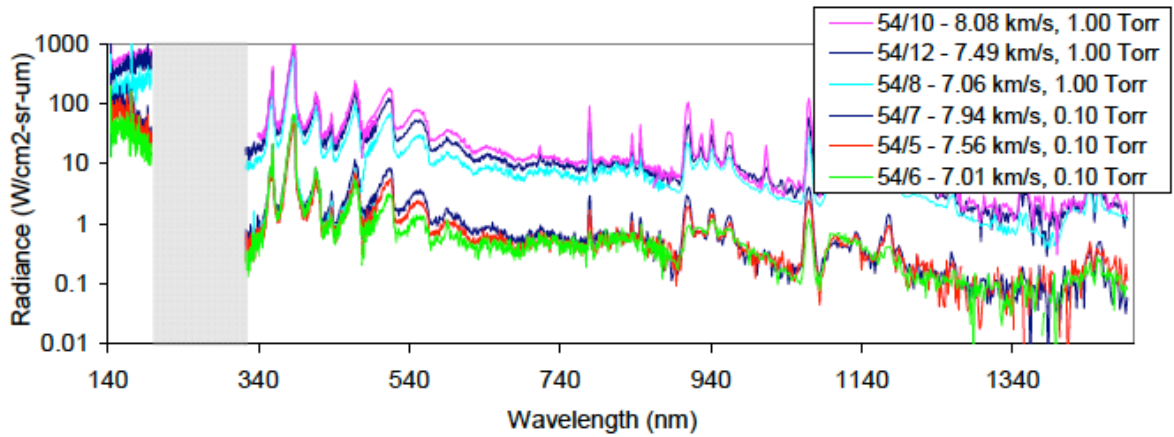

**Figure 30: Composite spectra for TC 2 nominal conditions (from [26])**

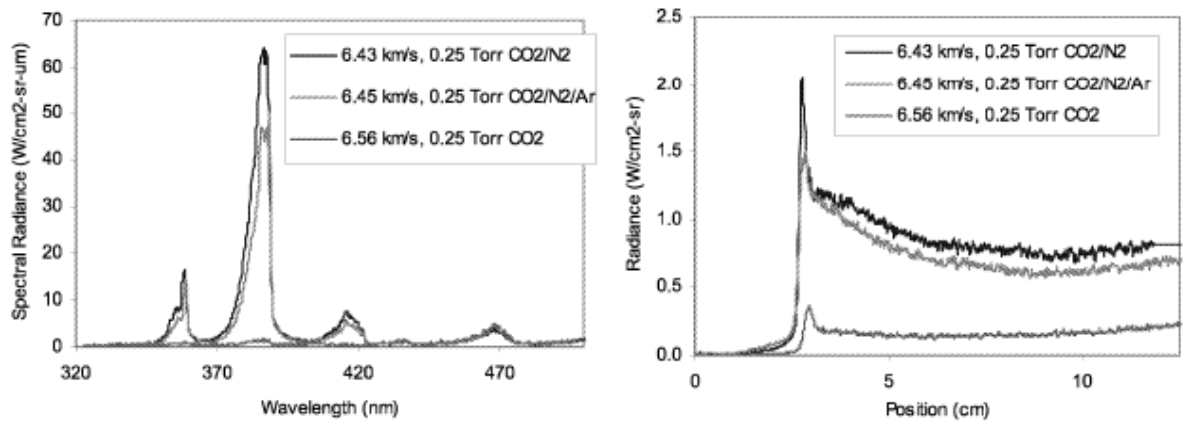

**Figure 31: EAST data for different mixtures: Left: Spectral radiation; Right: Radiation as function of position (from [29])**

More recently, Cruden et al [29] have investigated different Mars-like mixtures in order to assess the radiance in $CO_2/N_2/Ar$ systems. Shock tube tests were carried out for different mixtures: pure $CO_2$, 96:4 $CO_2/N_2$, and 95.8/2.7/1.5 $CO_2/N_2/Ar$ per mole. Tests were clustered around velocities and pressure of 6.5 km and 0.25 Torr, and 7 km/s and 0.1 Torr. Spectral range from VUV to mid IR has been covered and special care was given to CO (4+) and CN violet radiation. Measurements proved a low sensibility of radiation to the presence of $N_2$ except for CN Violet. Analysis of CN relaxation time has been performed using the CN Violet relaxation measurement, it has been found to be weakly dependent on the mixture composition. The experimental data were also compared against the predictions performed with HARA non-equilibrium radiation code [30] in [31]. Some results are shown in Figure 31, for the spectral radiation in the CN Violet and as function of position at 6.5 km/s and 0.25 Torr, this for the three mixtures. Figure 32 summarizes the contribution to the radiation of the different bands (VUV, UV, and VIS-IR) for the different mixtures as function of velocity.

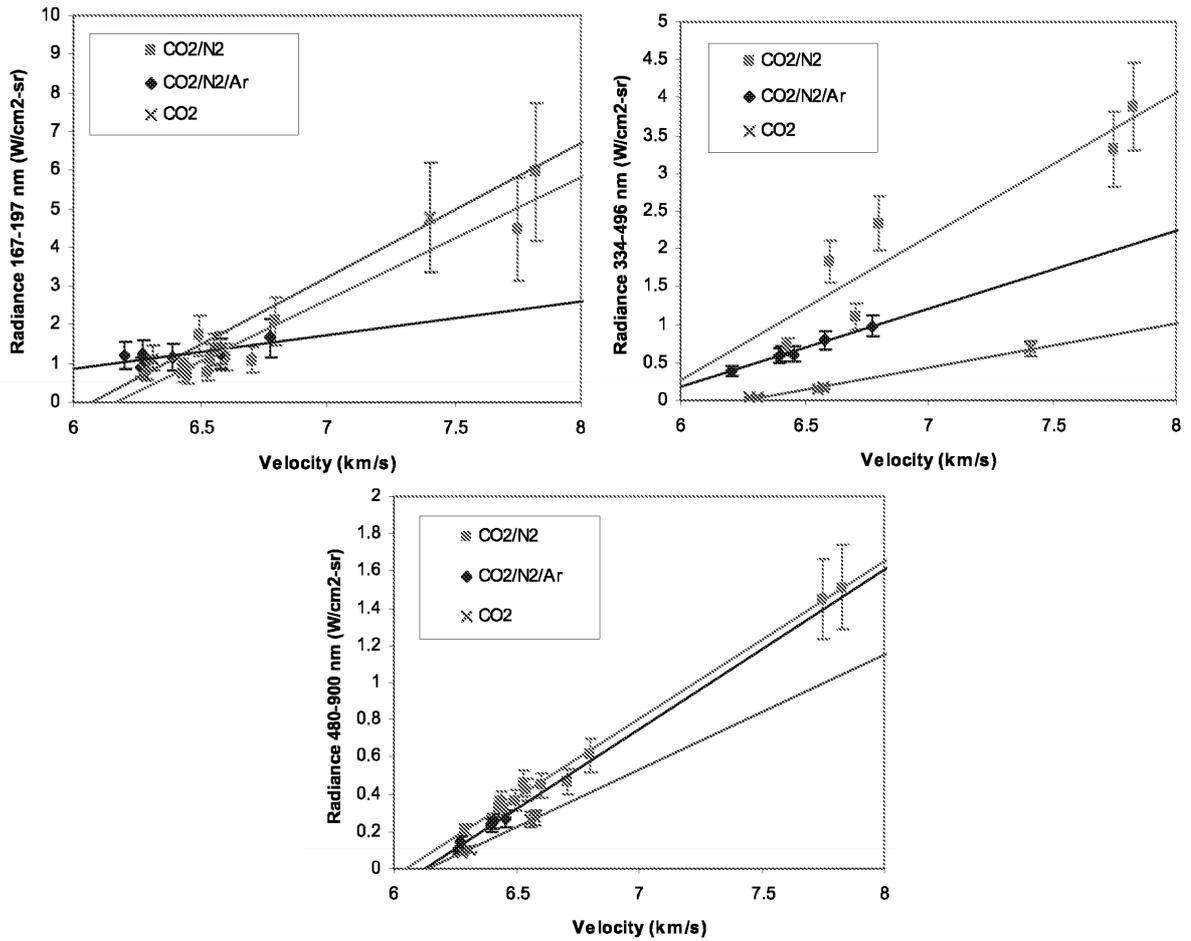

**Figure 32:** Steady state radiation for three Mars mixtures (pure $CO_2$, 96:4 CO2/N2, and 95.7:2.7:1.6 $CO_2/N_2$/Ar) at 0.25 Torr: Top left: 167-197 nm, Top right: 334-496 nm, Bottom: 480-900 nm (from [29])

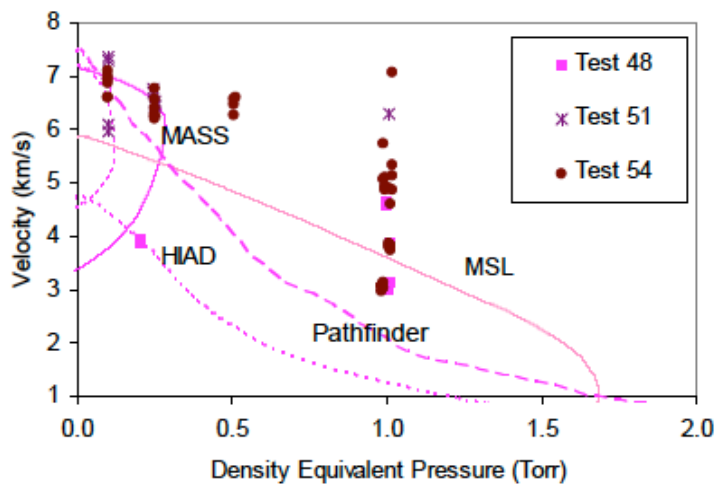

**Figure 33:** Summary of test conditions for which IR radiation was investigated in EAST with selected Mars entry trajectories (from [32])

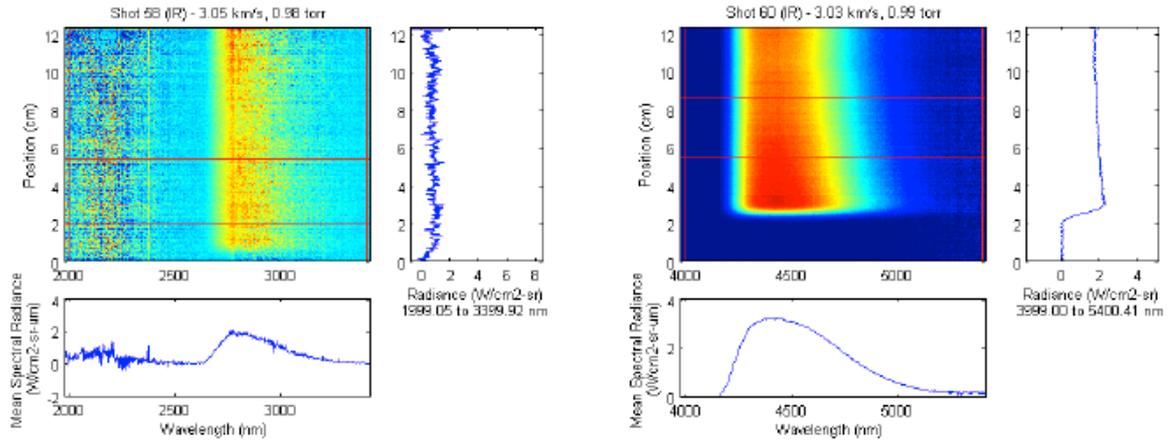

**Figure 34:** Spectral and spatial radiation measurements at 3 km/s and 1 Torr in pure $CO_2$ (left: 2700 nm band, right 4300 nm band) (from [32])

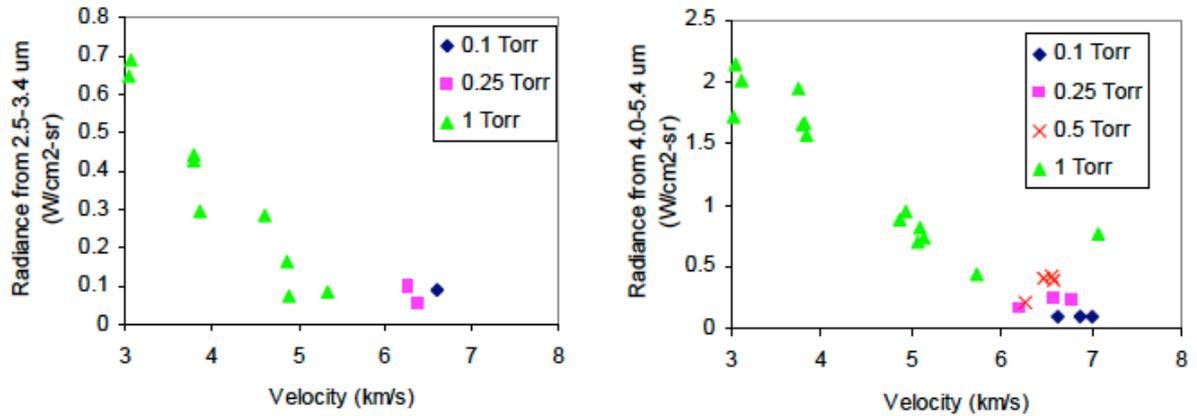

**Figure 35:** Radiation behind the shock for different velocities and pressures (left: 2700 nm band, right 4300 nm band) (from [32])

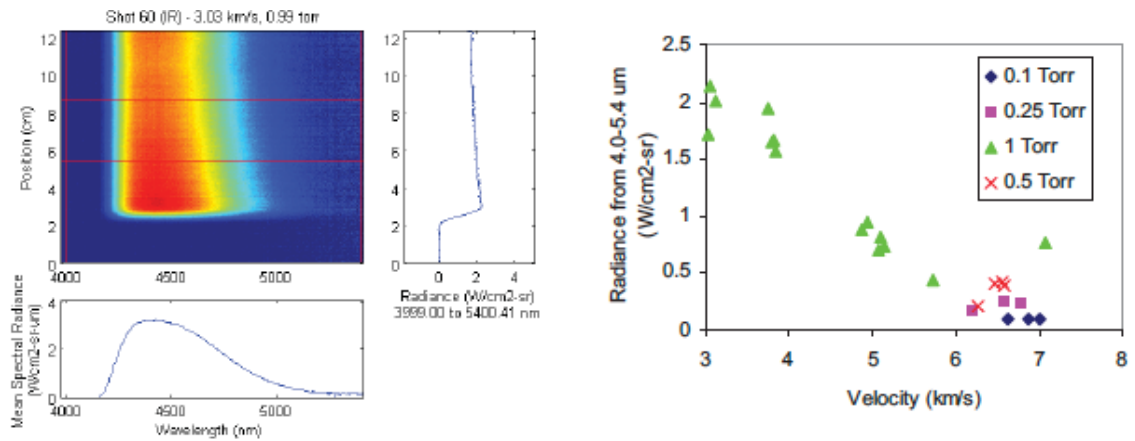

**Figure 36:** Left: Measurement of Mid-IR radiation for Mars entry conditions; Right: Total Mid-IR radiation as function of shock velocity (from [20])

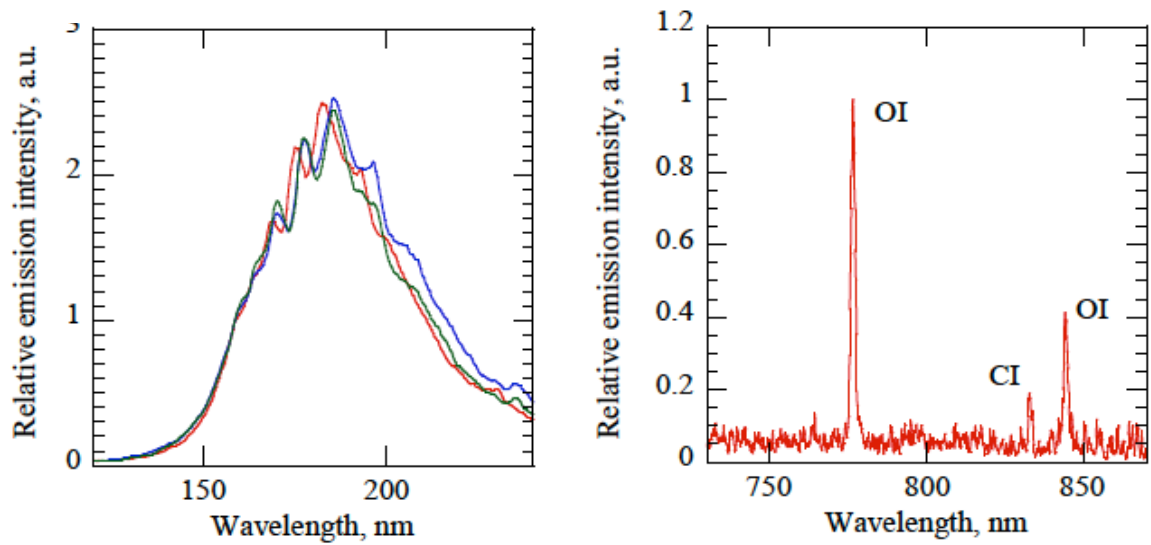

**Figure 37: Left: Emission spectrum in VUV for the different runs at 7.1 and 7.3 km/s. right: Emission spectrum in near region at 7.1 km/s and 200 Pa (from[33])**

Another recent contribution [32] has been more focused on mid-IR radiation that was investigated in EAST from 3 to 7.5 km/s and pressures from 0.2 to 1 Torr. Infrared radiation was found to be the highest at low velocity (3 km/s) and was decreasing with the velocity; the corresponding test conditions are summarized in Figure 33. Some data obtained for the 2700 nm and 4300 nm bands are reported hereafter. Figure 34 shows the spatial and spectral data at 3 km/s and 1 Torr in pure $CO_2$ for the 2700 and 4300 nm bands. Figure 35 summarized the results obtained for the IR bands in terms of radiation heating as function of the shock velocity.

Typical spectrum obtained for mid-wave IR radiation is shown in Figure 36. In the same figure the total radiation emitted in the mid-wave IR as function of the shock velocity is plotted.

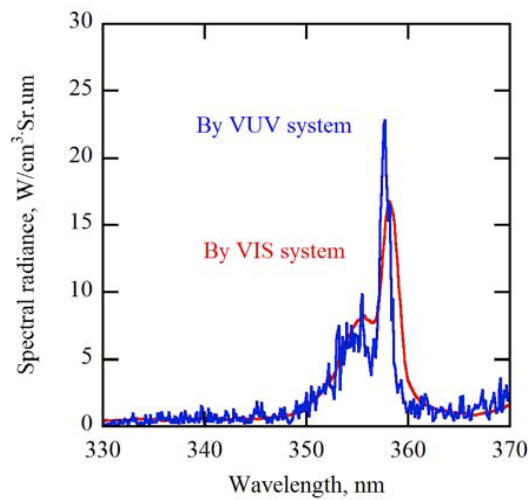

**Figure 38: Comparison of the measured CN violet bands in UV region obtained using VUV and VIS spectrometers (from [34])**

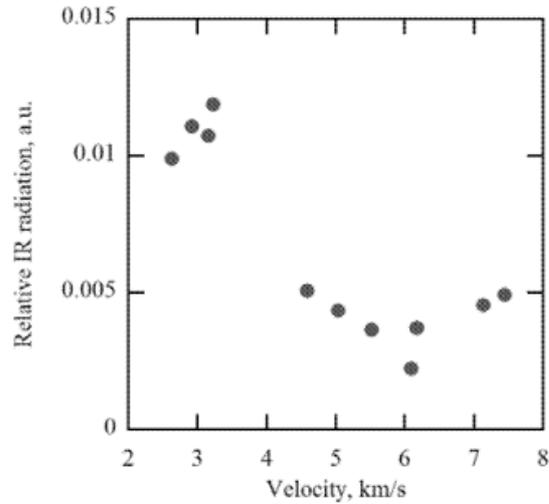

**Figure 39: Infrared radiation intensity as function of shock velocity measured in HVST (from [35])**

## 5.3 JAXA

Several test campaigns with $CO_2$ flows have been carried out in JAXA HVST shock-tube and HVET expansion tube [4]. First efforts [33-34-35] were dedicated to high velocity Mars entry, characteristic of a Mars aerocapture, later on lower velocities [36] were investigated. First results [33] were obtained for pure $CO_2$ at velocities of 7.1 and 7.3 km/s and pressure of 200 Pa. Radiation measurements were performed from VUV to NIR region, some of the data are plotted in Figure 37. In the same study, the $C_2$ Swan band was also investigated, and obtained data shown apparently an equilibrium state. In NIR region, atomic lines for carbon and oxygen were observed (see Figure 37).

Mars atmosphere with 96% of $CO_2$ and 4% of $N_2$ for high velocities was studied in [34]. Tests were performed at 7 km/s and 8.5 km/s for pressures of 1 and 0.1 Torr respectively. Radiation measurements were carried out from VUV to NIR focusing on CO(4+), CN Violet, and $C_2$ Swan systems using two spectrometers. Results obtained using the different measurement systems have been compared; corresponding results for CN violet are shown in **Figure 38**.

Infrared radiation has been extensively investigated in [35] in which experiments were conducted from 2.5 to 7.5 km/s and a pressure of 1 Torr. Measurements were focused on radiation intensity, this last was found to be high at low velocity and to decrease up to 6 km/s, after this minimum, radiation intensity was increasing with the velocity; this is summarized in Figure 39.

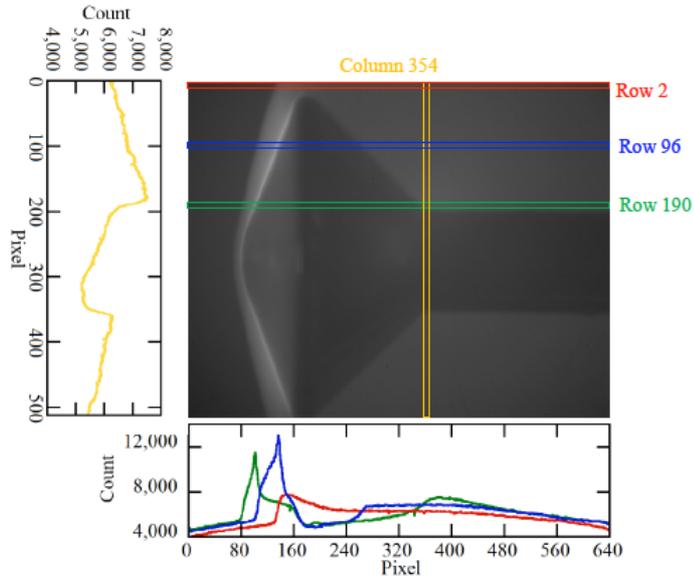

**Figure 40: Spatial distribution of IR radiation intensity at 7.6 km/s (from [36])**

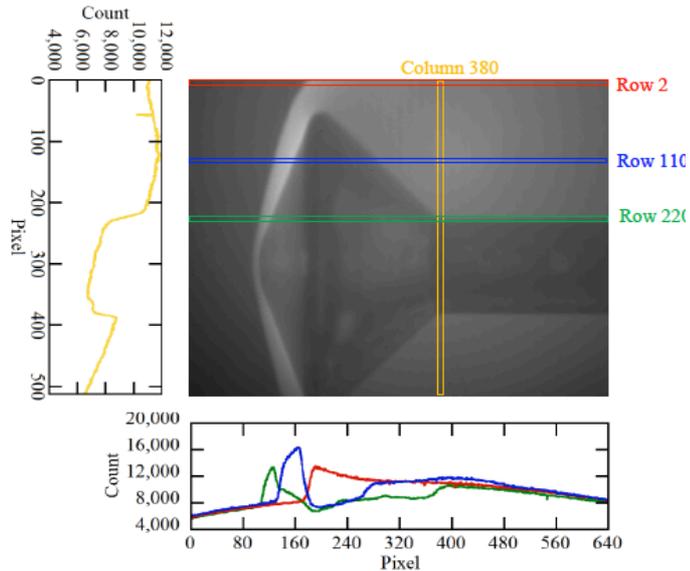

**Figure 41: Spatial distribution of IR radiation intensity at 5.5 km/s (from [36])**

The most recent test campaign [36] carried out by JAXA for $CO_2$ radiation has been performed in the HVET expansion tube. Tests were carried out for pure $CO_2$ at velocities of 7.6 and 5.5 km/s, and pressures of 0.075 and 0.15 Torr (10 and 20 Pa) respectively. A blunt body model was inserted within the flow during the experiments. Spatial distribution of VIS and IR radiation was studied using high-speed imaging systems; corresponding results are shown in Figures 40 and 41. IR radiation spectra were also obtained with and without the model, however, according of the authors [33], effects of chemical freezing in the expansion tube require a more comprehensive investigation for a more refined assessment of the results.

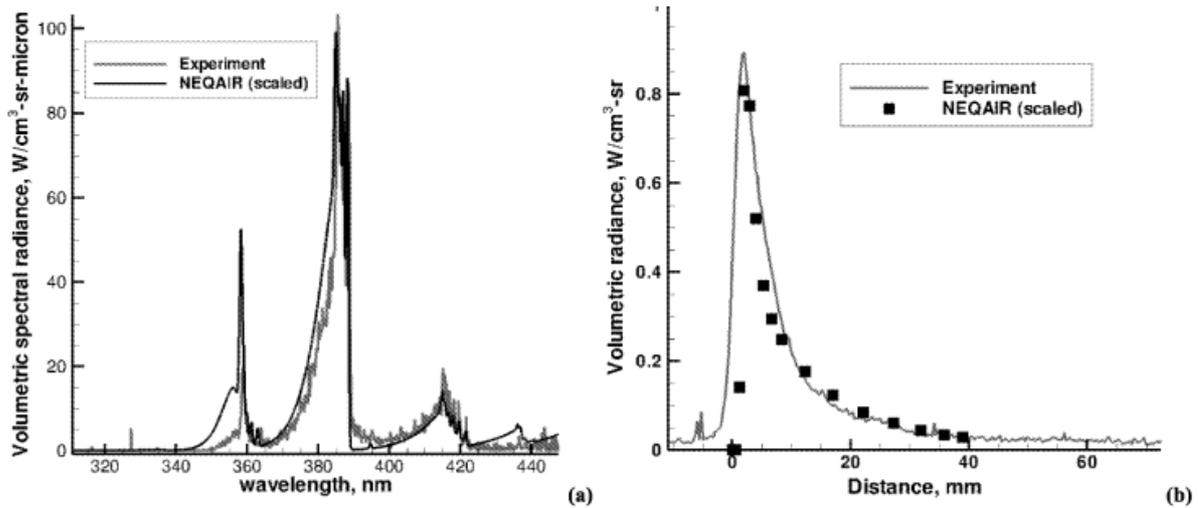

**Figure 42: Comparison of NEQAIR and X2 test data: a) volumetric spectral radiance between 310 and 450 nm; b) volumetric radiance along stagnation line (from [37])**

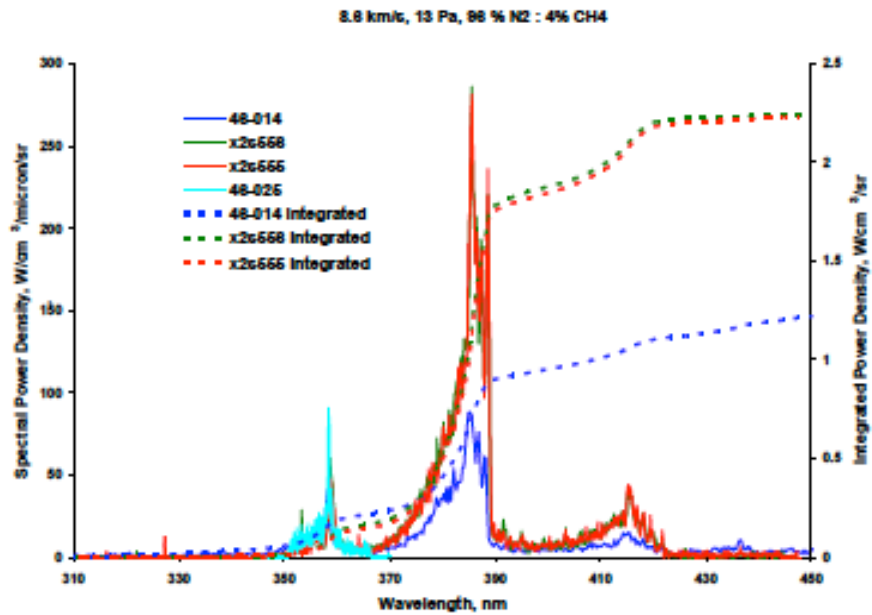

**Figure 43: Wavelength comparison between EAST and X2 for Mars conditions (datasets from X2 are started by x2 in the legend) (from [37])**

## 5.4 X2

Different test campaigns [37-38-39-40] have been carried out at the University of Queensland, in the X2 expansion tube, dedicated to radiation measurements into Mars mixture atmosphere. In [37-38] tests have been performed for a Mars atmosphere (96% $CO_2$, 4% $N_2$), at a shock speed of 8.6 km/s, and a free stream pressure of 0.1 Torr. Test data have been numerically reconstructed using NEQAIR and comparisons between numerical predictions and experimental data can be seen in Figure 42. Radiation measurements were focused on CN violet band system.

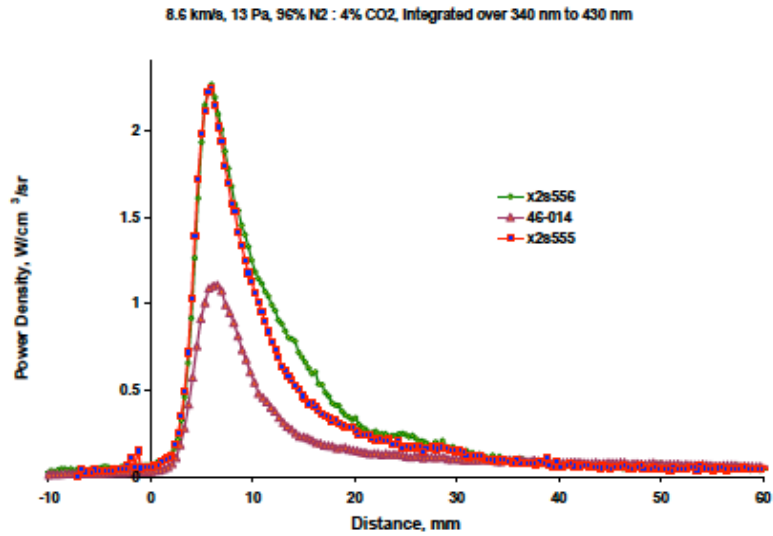

**Figure 44: Power density comparison between EAST and X2 for Mars conditions and 0.1 Torr (datasets from X2 are those with the name starts by x2 in the legend) (from [37])**

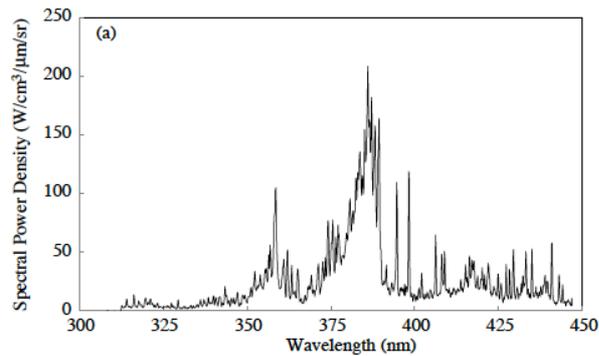

**Figure 45: Spectral power density for Mars gas expansion tube study (from [40])**

| Velocity (km/s) | 8.5 |
| --- | --- |
| Density (kg/m$^3$) | 1.79 10$^{-3}$ |
| Temperature (K) | 1060 |
| Enthalpy (MJ/kg) | 36 |
| Mach number | 12.8 |

**Table 2: Test in X2 for a Mars-like mixture (96 % $CO_2$, 4% $N_2$): Calculated free stream conditions in [40]**

Experimental results obtained for close conditions (same pressure and shock velocity of 8.7 km/s) have been compared with EAST data [37]. This is displayed in Figure 43 for the spectral wavelength radiance and in Figure 44 for the power density.

Other similar conditions with a shock velocity higher than 8 km/s have been investigated in [40]. Test conditions are reported in Table 2 and corresponding results in Figure 45. Measurements were again concentrated on the CN violet system.

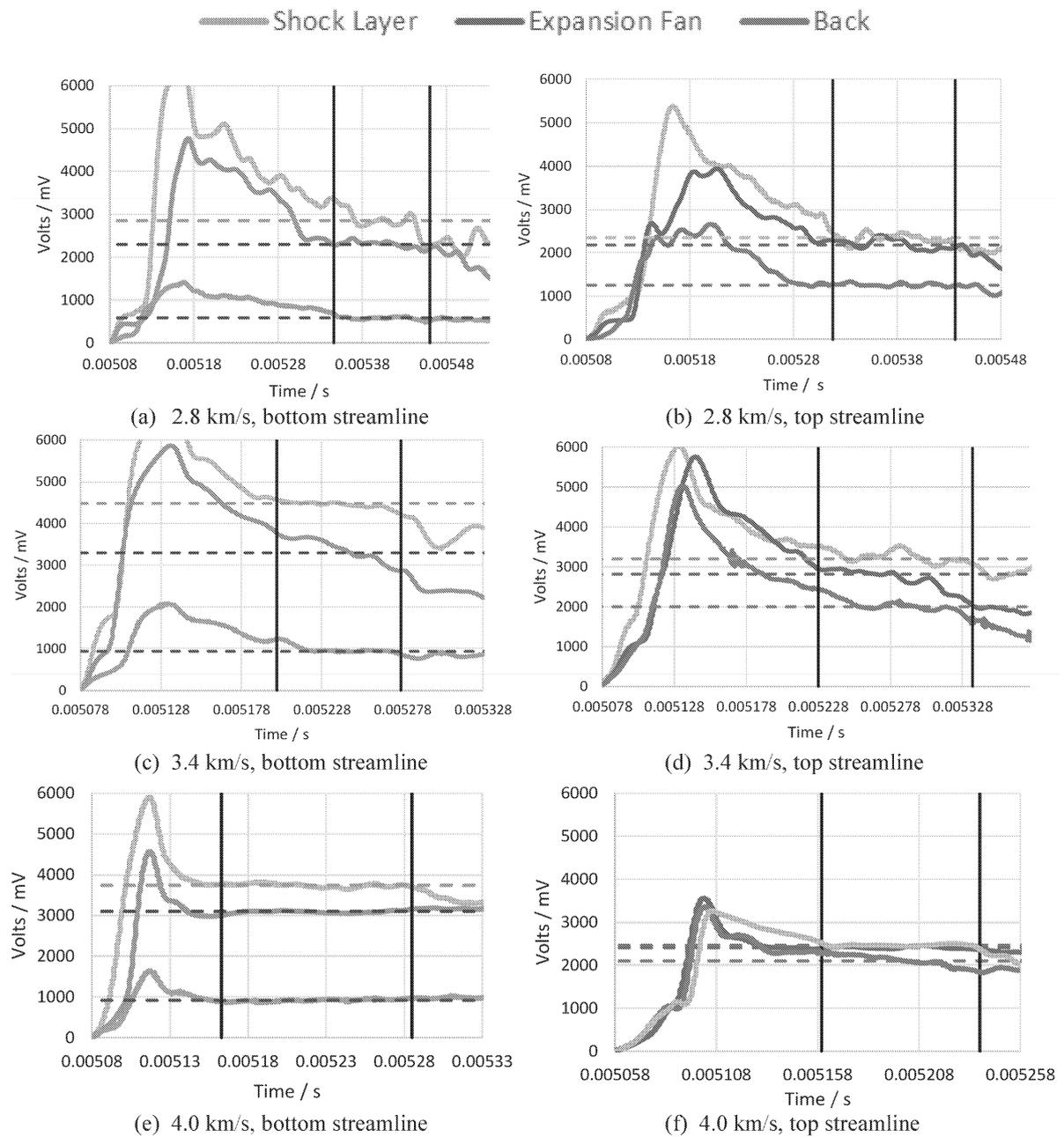

**Figure 46: Time-dependent radiation measurements (from [39])**

In [39] tests have been performed at freestream velocities of 2.8, 3.4, and 4 km/s, and spectroscopy measurements focused on $CO_2$ radiation at 4300 nm. The test campaign was performed for supporting the NASA MSL mission. The objective was to investigate the afterbody radiating expanding flow, since afterbody radiation is an issue at these velocities. As a consequence, pure $CO_2$ was used for the test, since afterbody radiation is mostly produced by $CO_2$. Test conditions are summarized in Table 3, temporally resolved radiation measurements in Figure 46. Comparisons were performed against numerical calculations performed with NEQAIR for CO band radiance.

| V (m/s) | P (Torr) | T (K) | c (m/s) | Tv (K) | $CO_2$ (mole fraction) |
|---|---|---|---|---|---|
| 2877 | 2.7 | 1191 | 507 | 2358 | 0.92 |
| 3484 | 2.7 | 1378 | 560 | 2758 | 0.76 |
| 4077 | 1.13 | 1281 | 543 | 2815 | 0.73 |

Table 3: Estimated X2 freestream conditions for MSL project [39]

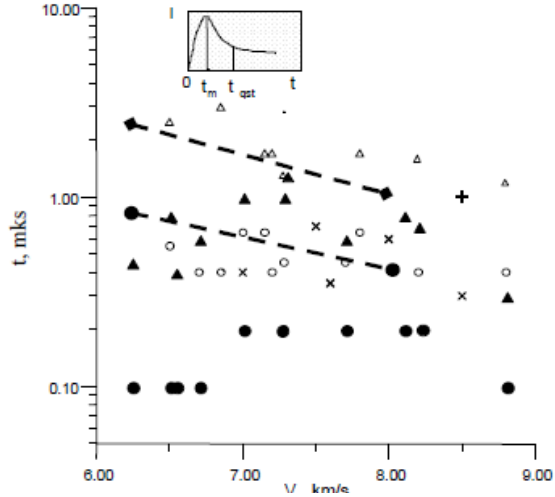

Figure 47: Measured and calculated time-response characteristics of the non-equilibrium radiation peak CO(4+) and CN(violet) behind shock wave vs velocity. 1)measurements ● – $t_m$ (time to reach radiation maximum during a shot) for CO(4+); ▲ – $t_{qst}$ () for CO(4+); Δ – $t_{qst}$ (characteristic time for non-equilibrium radiation peak duration) for CN(violet); ○ – $t_m$ for CN(violet); X from [41]. 2) calculations: curves – $t_m$ for CN(violet), – $t_{qst}$ for CN(violet) (from [42])

### 5.5 Data from other facilities

#### 5.5.1 ADST

ADST is a high enthalpy shock-tube [4] located at TSAGI near Moscow. Some tests related to Mars entry were performed in the frame of ISTC project [42]. A test campaign was performed for the following conditions:

- Pressure of 0.2 Torr;
- Mixture of 70 % [$CO_2$] and 30 % [$N_2$];
- Velocity from 6 up to 9 km/s.

Measurements were performed for the absolute intensity of the main radiative components such as CO(4+) and CN(violet). Figure 47 shows the time-response characteristics of the non-equilibrium radiation peak as function of velocity, for CO(4+) radiation at a wavelength of 200 nm and the bands system CN (violet) radiation measured in ADST. In this figure, $t$m denotes the time for which the maximum radiation is reached, while $t_{qst}$ is the characteristic non-equilibrium radiation peak duration), this figure shows also similar measurements taken from [41]. The peak radiation intensity of CO(4+) measured in ADST compared against calculated values is plotted as function of shock velocity in Figure 48.

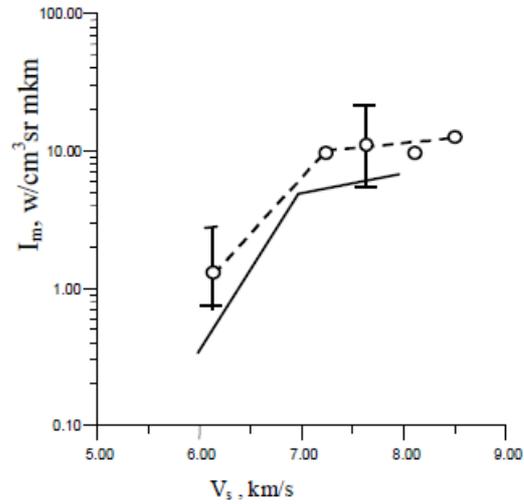

**Figure 48: Peak values of radiation intensity ($I_m$) of CO(4+) behind the shock-wave as function of velocity: Curve – calculation, ○ – .measured radiation intensity (from[42])**

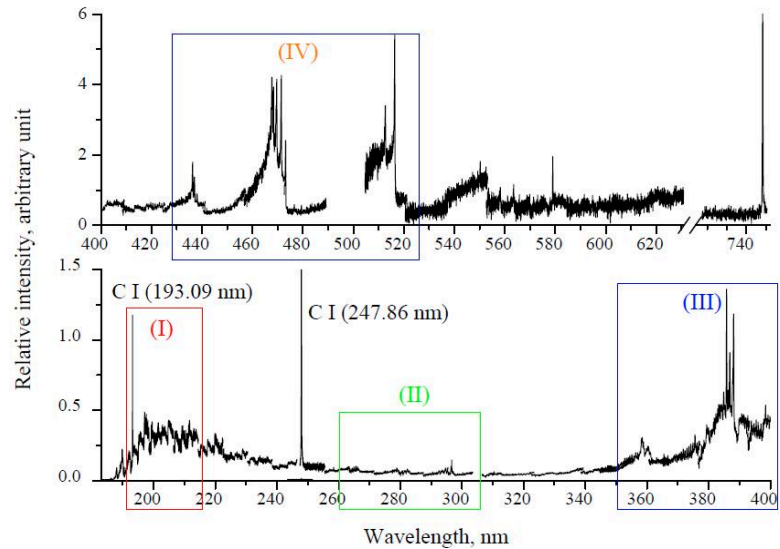

**Figure 49: Spectrum of pure $CO_2$ from VUV to VIS in ICP CORIA torch (from [43])**

### 5.5.2 CORIA ICP torch

At least two test campaigns with pure $CO_2$ have been performed in the CORIA ICP torch for subsonic flow conditions [43-44]. Pressure and specific test enthalpy were respectively of 38 mbar and 25 MJ/kg in [43] and 90 mbar and 8.5 MJ/kg in [44], More details on the measurements are available on this last study. Observed spectrum for the first test conditions [43] is reported in Figure 49. The spectrum can be divided in 4 parts: In the VUV range (part I), the contribution of CO(4+) is well identified, while in part II, the CO(3+) system is present. $C_2$ Deslandres-d'Azambuja system is shown in part III, and $C_2$ Swan in part IV. Some atomic lines for carbon can be also identified, as well as some radiation from CN. This last is present due to the introduction of air during the test for pressure control purpose.

### 5.5.3 ICARE PHEDRA facility

PHEDRA ICARE facility (former SR5 at CNRS Orléans in France) is an arc-jet dedicated to the investigation of low-density flows. Several studies [45-46] have been conducted to investigate Mars plasma flows.

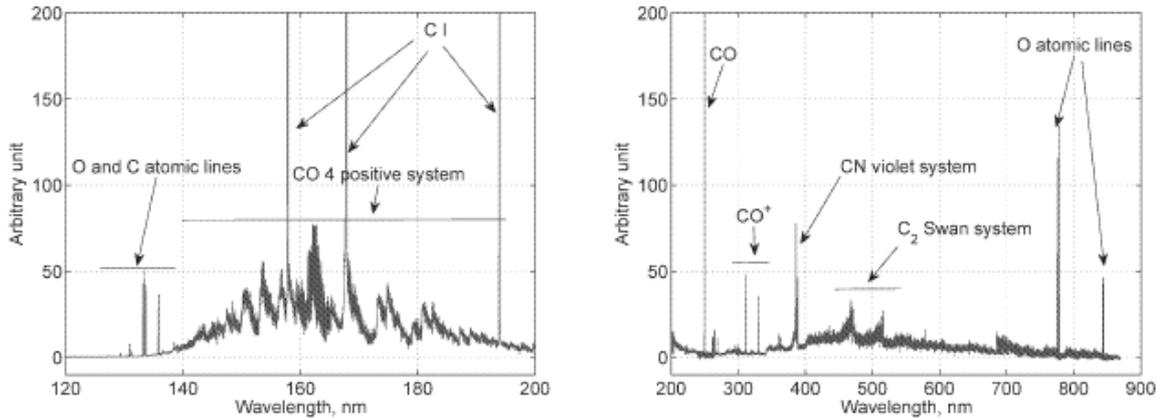

**Figure 50: Radiation systems and lines in PHEDRA (from [46])**

Recently, a test campaign [46] has been conducted in this ground test for measuring Mars plasma radiation from 110 nm to 900 nm. Tests have been performed with a 97% $CO_2$ – 3% $N_2$ mixture, at enthalpies in between 4.16 and 35 MJ/kg. Chamber static pressure was in between 2.9 and 5.3 Pa. The emission spectra obtained are shown in Figure 50, in the left part for the VUV region, in the right part for the UV, VIS, and NIR. CO(4+) system, $C_2$ Swan and CN Violet systems can be identified, as well as molecular lines for $CO^+$ and CO, and atomic lines for O and C. However, the conditions, in term of flow enthalpy and pressure, for which these spectra have been obtained, are not documented in [46].

### 5.5.4 KAIST

KAIST shock-tube [4,47], located in South-Korea, can be used in reflected mode, a velocity of 3.45 km/s producing a reflected shock temperature corresponding to a Mars entry velocity of 6.4 km/s can be obtained for $N_2$ – CO mixture as achieved in [47]. This shock-tube has been used for radiation studies in $N_2$ – CO mixture [47]. The wavelength region observed in these experiments covers the entire O–O band of the CN violet emission.

The chemical process of CN formation in a CO-$N_2$ mixture with 78% of CO and 22% of $N_2$, has been studied through temporally resolved intensity measurement of CN violet radiation in the reflected shock region of the shock tube [47]. The test conditions are reported in Table 4. $P_{eq}$ and $V_{eq}$ denoted the equivalent pressure and velocity obtained in reflected mode.

| Run | Pressure (Torr) | Velocity (km/s) | $P_{eq}$ (Torr) | $V_{eq}$ (km/s) |
|---|---|---|---|---|
| 1 | 1.3 | 3.45 | 3.77 | 5.2 |
| 2 | 2.6 | 3.2 | 7.3 | 4.83 |
| 3 | 2.6 | 3.4 | 8.37 | 5.13 |

**Table 4: Test conditions run in KAIST shock-tube [47]**

The measured radiation emission intensity and pressure behind the reflected shock are shown in the left part of Figure 51. In the right part of the figure the experimental results are compared with the calculations obtained with several chemical kinetics models for the three different cases. As shown in the figure, the emission intensity reaches a near-equilibrium value at 15 ms for all cases. The models overpredict the intensity at the non-equilibrium peak for all cases.

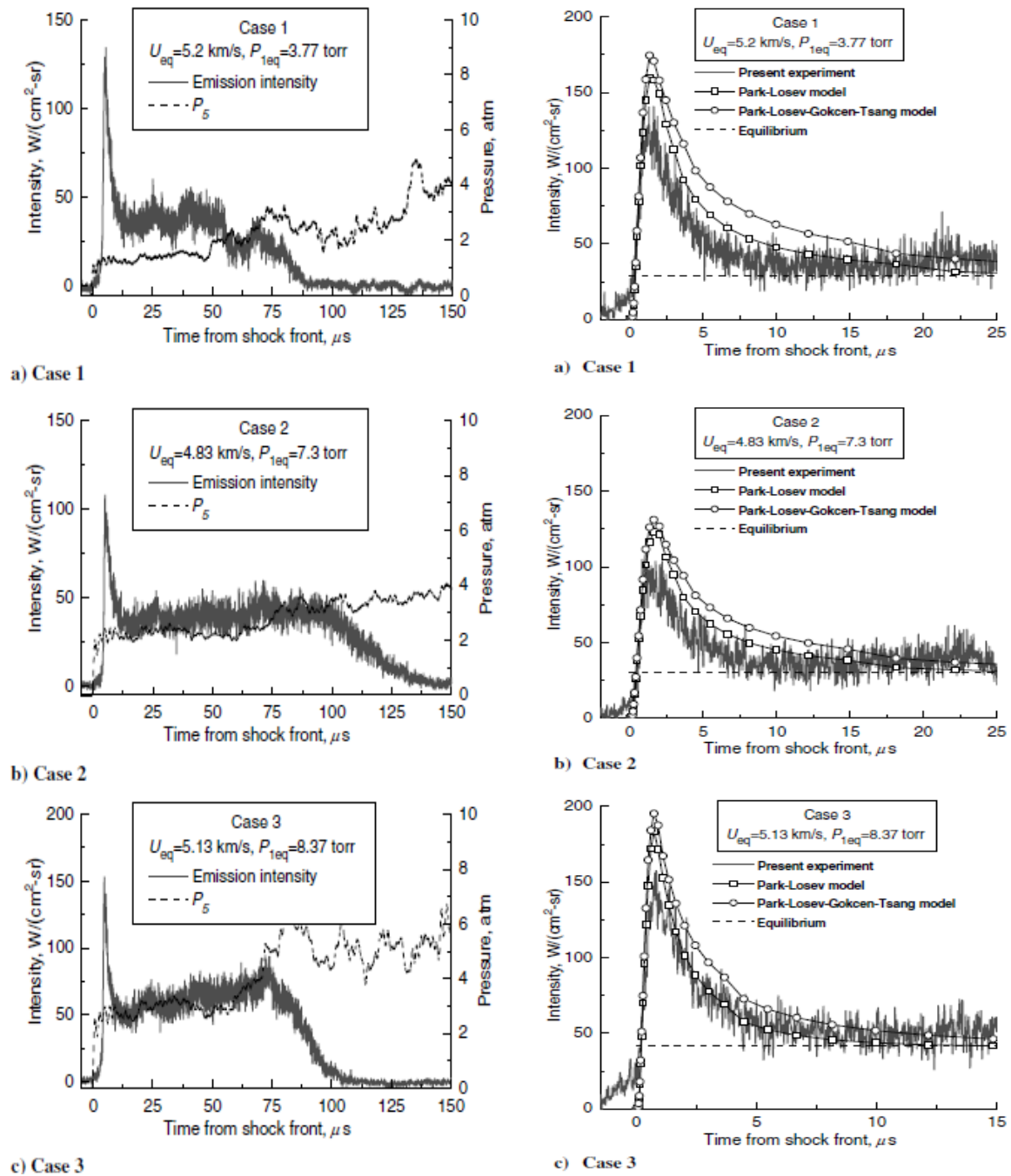

**Figure 51: Left: Measured intensity and pressure behind the reflected shock. Right: Comparisons of numerical results with measurements (from [47])**

### 5.5.5 LAEPT ICP torch

Several test campaigns [48-49] have been conducted in the LAEPT ICP torch at the University of Clermont-Ferrand (France) for investigating $CO_2$ plasma emission. Experimental data have been obtained for a pressure of 950 hPa and a 97%-3% $CO_2$-$N_2$ mixture. A temperature of 6000 K was estimated on the plasma jet axis using numerical fitting of emission spectra with the line-by-line code SPARTAN [50]. Emission spectra are shown in Figure 52, CN Violet emission is predominant while $C_2$ emission is very weak, radiation from CN Red, NO and $N_2^+$(1-) is not observed. Atomic lines from C and O are present.

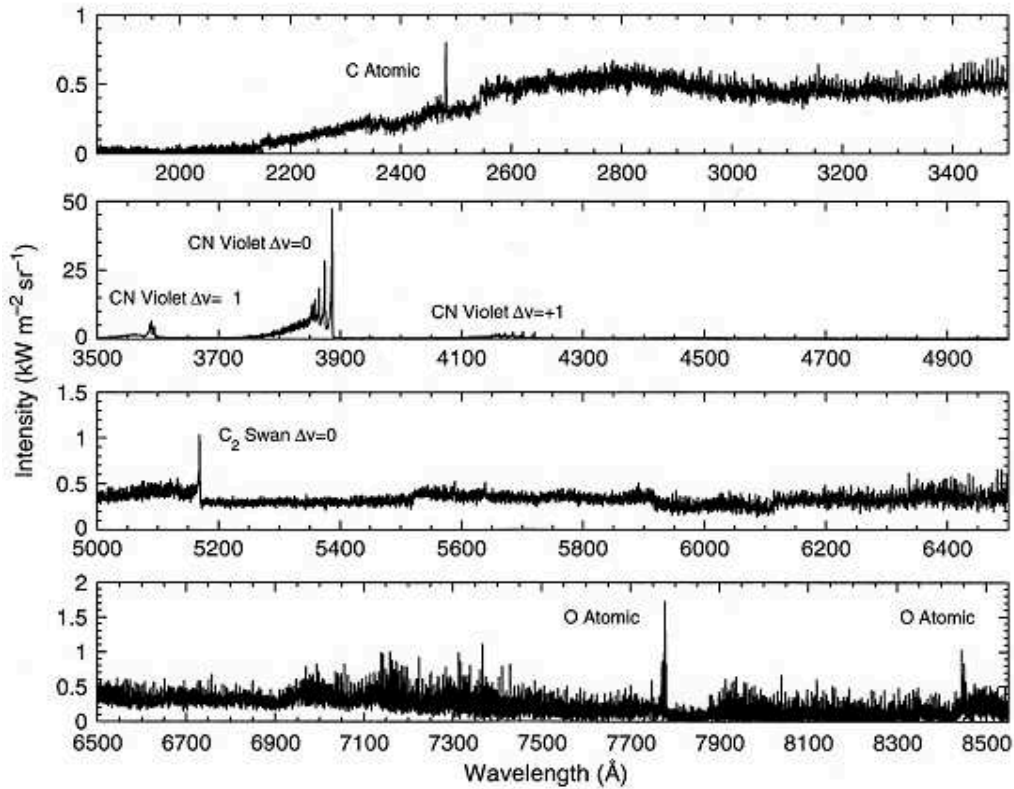

**Figure 52: Emission spectrum of a $CO_2$ plasma in the LAEPT ICP torch (from [48])**

### 5.5.6 LENS

The LENS-XX expansion tube has a large measurement equipment [51] and has already provided valuable radiation data for Earth high-speed re-entry. For the NASA CEV project a large number of shots have been performed for Lunar return conditions [52] and data were collected from VUV to near infrared. The facility has been also used for investigating $CO_2$ flows around a blunt-body [53] for laminar and turbulent conditions, with and without dissociation and radiation but no spectral data was found to be available in the literature in relation with this study.

### 5.5.7 Moscow State University

A double diaphragm shock-tube is available at the Moscow State University (MSU). $CO_2$-$N_2$ mixture (70%-30%) at high velocity has been investigated by Kozlov et al [54]. Two test conditions have been studied, at shock speeds of 4.6-5.8 km/s and 1 Torr (133 Pa), and 6.3-7.6 km/s and 0.3 Torr (40 Pa). Spectroscopic measurements were focused on 200-850 nm wavelength range. For both conditions, molecular radiation of CN Violet, NO, and CO was observed. $C_2$ Swan emission was only observed for velocities higher than 6.3 km/s, and atomic oxygen triplet only at shock speeds higher than 6 km/s. Figure 53 displays the experimental data obtained at 0.3 Torr, this highlights the presence of CO(4+) and $N_2$(2+) systems at the high velocities. Some of the tests have been numerically reconstructed in [5], the predictions were able to reproduce the experimental trends and were in the experimental margins of accuracy.

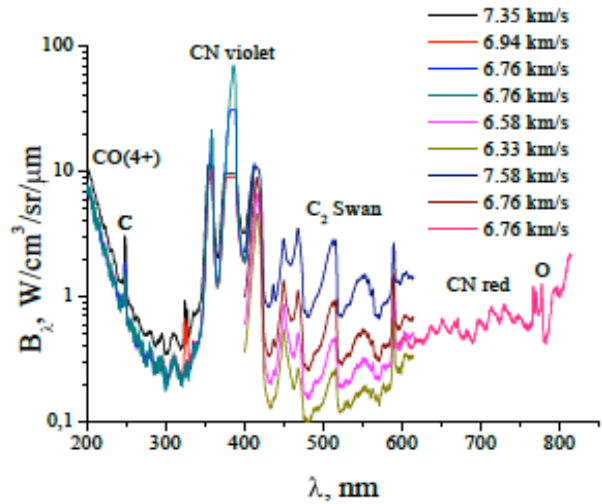

**Figure 53: Spectral power densities for a pressure of 0.3 Torr (from [54])**

### 5.5.8 TCM2

Test campaigns related to Mars entry have been conducted in TCM2 in the frame of ESA TRP activities [55-56], and of the Ph. D. thesis of D. Ramjaun [57]. Some of the conditions are reported in Table 5. This facility has been extensively used for radiation and thermo-chemistry studies related to Earth, Mars and Titan entries. It has provided the major part of the results produced in Western Europe on these topics until its closure.

| Case | Velocity (km/s) | Pressure (Torr) | Mixture |
|------|-----------------|-----------------|---------|
| 1 | 5.8 | 1.47 | 30% [$N_2$], 70% [$CO_2$] |
| 2 | 5.2 | 4.06 | 100% [CO] |
| 3 | 6.17 | 1.8 | 70% [$N_2$], 30% [$CO_2$] |
| 4 | 6.25 | 1.68 | 70% [$N_2$], 30% [$CO_2$] |

**Table 5: Tests carried out in TCM2 for Mars entry**

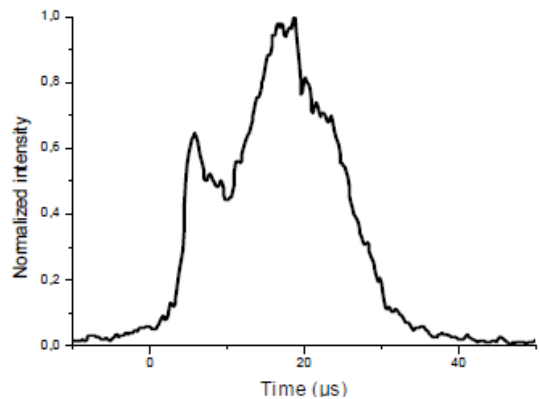

**Figure 54: Time profile of CN violet $\Delta v=0$ for Case 3 of Table 5 (from [55])**

Tests carried out in the frame of [55] cover a pressure range from 0.75 up to 82.7 Torr (100 up to 1100 Pa), and a velocity range from 4.3 up to 6.25 km/s. Available data with radiation spectra of CN Violet and CO Angström systems, temperature measurements were obtained in [55] for the Cases 3 and 4 of Table 5. The objective of this study was to investigate CN violet system that was expected to be the main radiator in the near UV and visible wavelength range (although CO(4+) is the largest radiator overall) and it was observed for its Δv=0 and Δv=-1 bands. CO Angström was also expected to take a non-negligible part of the radiation emission.

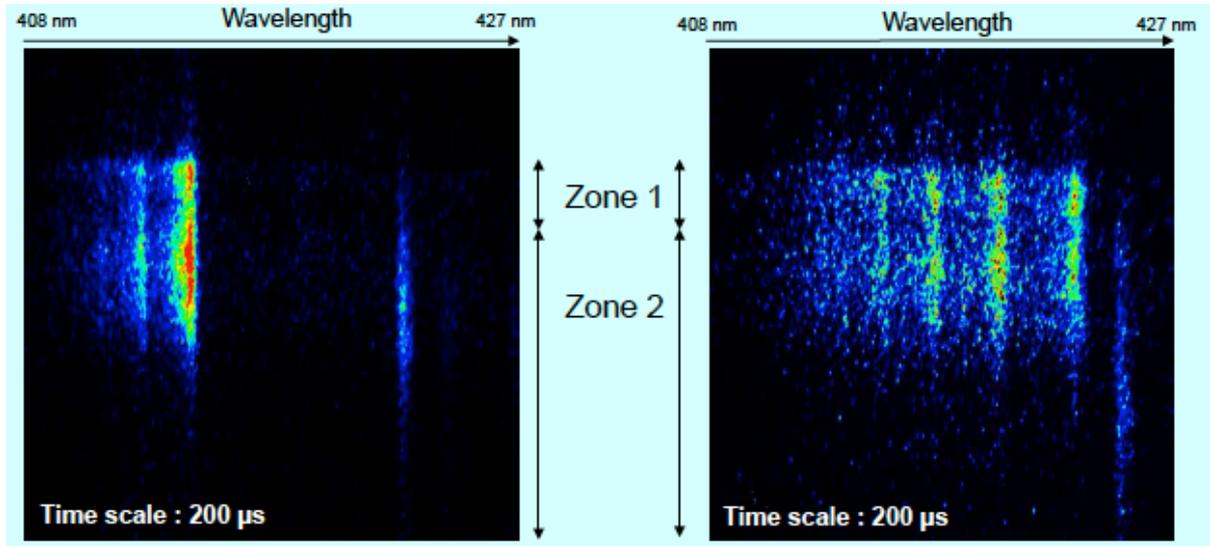

**Figure 55: Zones 1 and 2 considered in TCM2 for radiation study (from [55])**

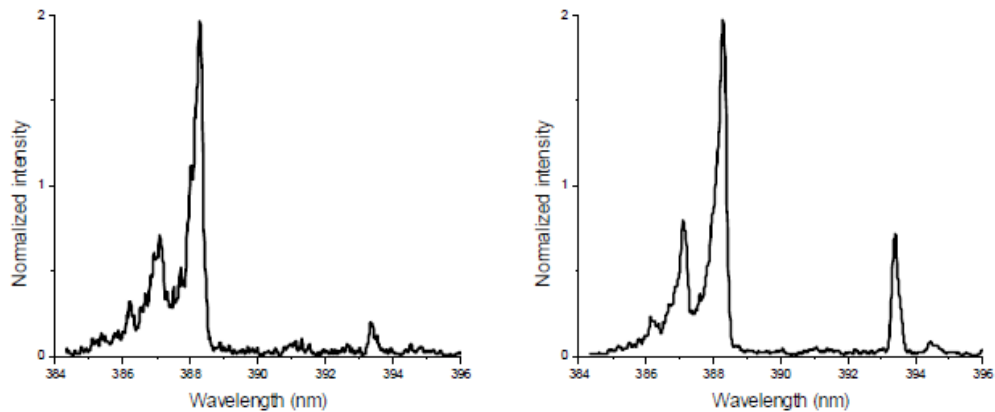

**Figure 56: Spectrum of CN violet Δv=0 (Case 3 of Table 5): Left: zone 1; Right: zone 2 (from [55])**

Time profile for CN violet is plotted in Figure 54, while the spectra of radiation intensity as function of wavelength for Δv=0 and Δv=-1 bands are shown in Figure 56 and Figure 57 respectively. Two areas (Zones 1 and 2 as shown in Figure 55) had to be taken into account separately for the spectral study. The temperatures are different between these two zones due to the difference in radiation intensity. For Case 3 (of Table 5), in Zone 1, vibrational and rotational temperatures were 2100 and 6000 K respectively, while their values are 2500 and 5500 K in Zone 2 [55]. The use of a broadband spectrometer showed that, the CO Angström emission was very weak compared to CN violet emission in the experimental conditions retained for the study. Recording CO Angström emission was made by sending more light in the streak system and then by decreasing the spectral resolution of the measurements. According to [55], these reasons explain why the spectra exploitation was difficult. But the time evolution extracted from the image shows, as for CN violet, two successive excitations.

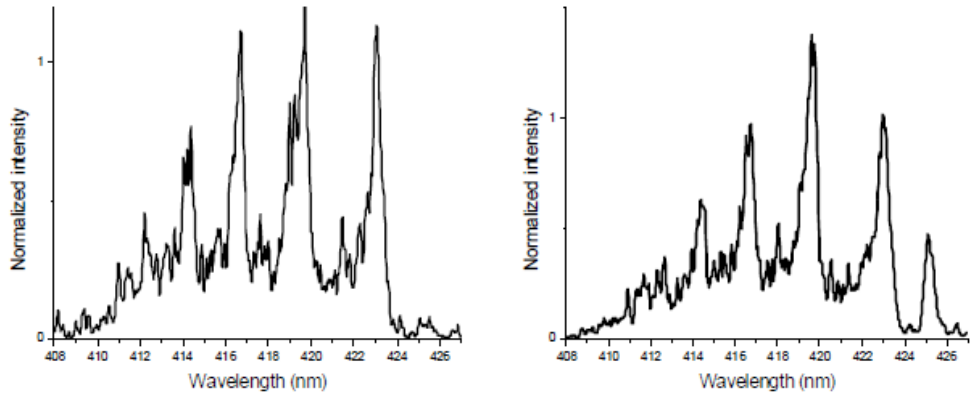

**Figure 57: Spectrum of CN violet Δv=-1 (Case 4 of Table 5): Left: Zone 1; Right: Zone 2 (from [55])**

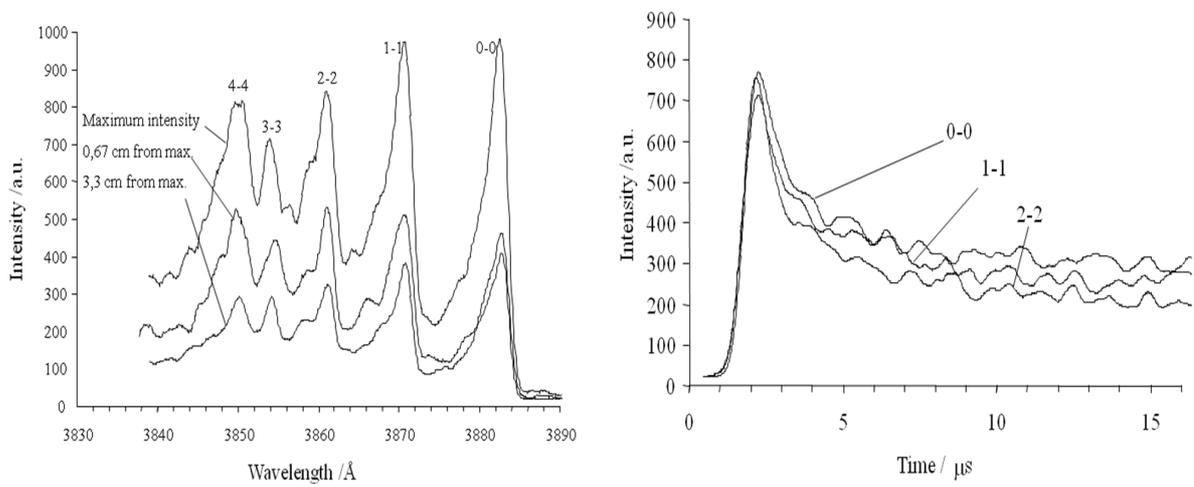

**Figure 58: Left: Spectrum of CN violet Δv=0. Right: Time profile of CN violet Δv=0 for the Case 1of Table 5 (from [57])**

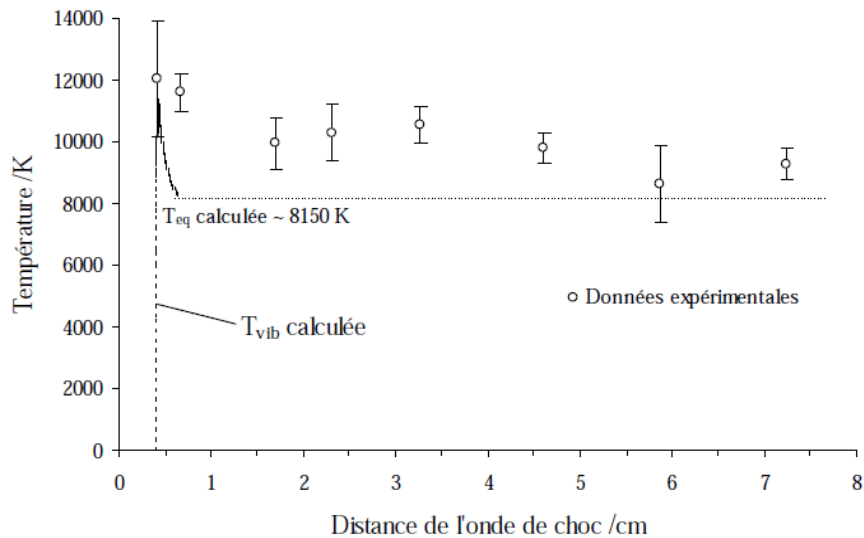

**Figure 59: Experimental and predicted vibrational temperature for Case 1 of Table 5 as function of shock distance (from [57])**

For this study [55], time average temperatures were measured. If rotational temperatures were high enough (about 6000 K), vibrational temperatures were only estimated at about 2500 K. This low value is in agreement with the theory, since $CO_2$ dissociation requires a large amount of energy, which is not used for the vibrational excitation.

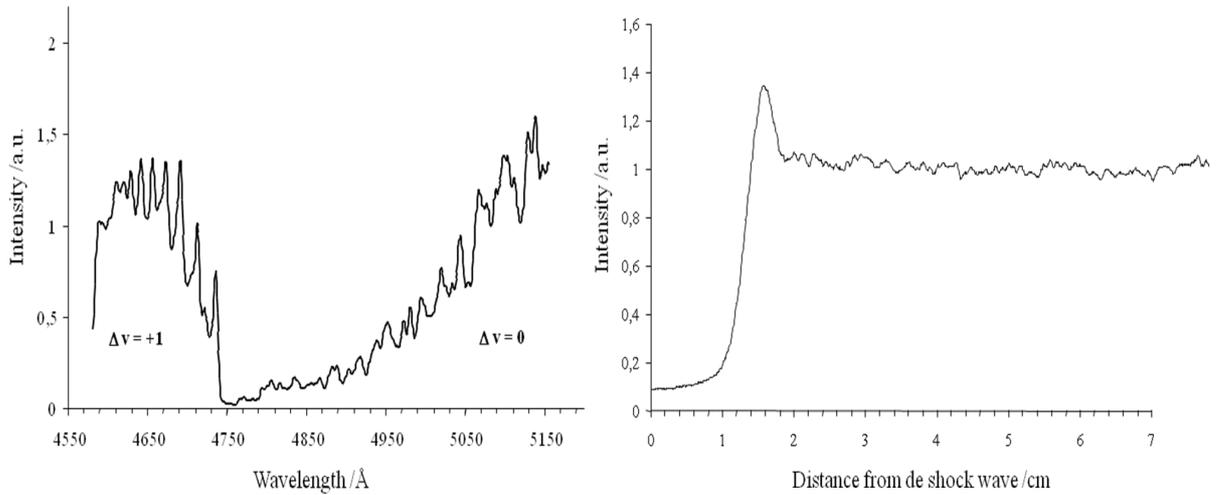

**Figure 60: Left: Spectrum of $C_2$ for Swan bands ($\Delta v=0$ and $\Delta v=+1$) for Case 2 of table 5 – Right: Time profile of $C_2$ for Swan bands ($\Delta v=0$ and $\Delta v=+1$) for same conditions (from [57])**

| Case | Velocity (km/s) | Pressure (Torr) | Mixture |
|---|---|---|---|
| 1 | 6.27 | 1.59 | 30% [$N_2$], 70% [$CO_2$] |
| 2 | 6 | 1.42 | 48.5% [$CO_2$], 1.5 % [$N_2$], 50% [Ar] |
| 3 | 7.49 | 0.1 | 0.3% [$N_2$], 9.7% [$CO_2$] 90% [Ar] |
| 4 | 7.06 | 0.6 | 3% [$N_2$], 97% [$CO_2$] |
| 5 | 6.5 | 1.32 | 30% [$N_2$], 70% [CO] |

**Table 6: Tests carried out in VUT-1 for Mars entry**

For Case 1 of Table 5, spectra of CN violet and time profile of emission intensity [57], shown in Figure 58 are available. The vibrational temperature has been also measured for Case 1 of Table 5 [57] and corresponding results are plotted in Figure 59. For the Case 2 with pure CO the objective was to investigate the $C_2$ Swan bands. Available spectra and time profile of emission intensity are shown in Figure 60.

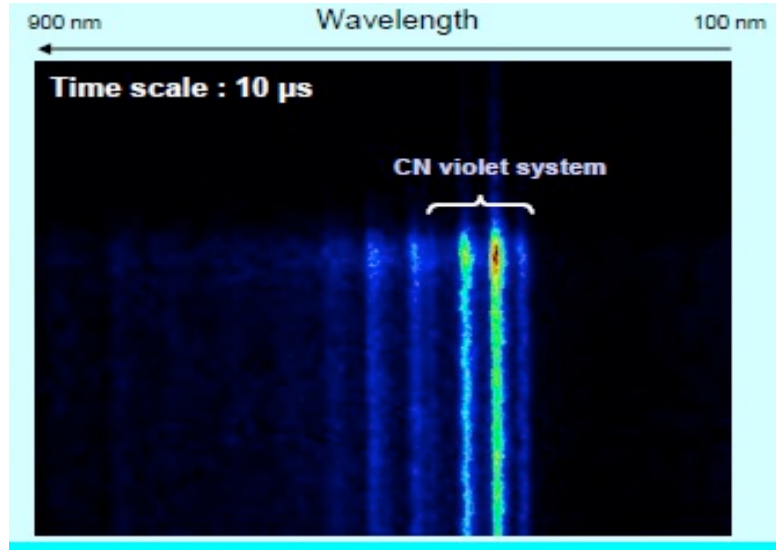

**Figure 61: Streak image for Case 1 of Table 6 (from [55])**

### 5.5.9 MIPT VUT-1

In the frame of ESA TRP activities, test campaigns have been carried out in VUT-1 at MIPT [55, 58] for Mars-like atmosphere. Other campaigns were conducted in the same facility in the frame of the ISTC programme [59]. For ESA activities, tests were conducted for the conditions resumed in Table 6. Other mixtures have been investigated they are listed hereafter:

- 90% Ar, 9.7% $CO_2$, 0.3% $N_2$;
- 50% Ar, 48.5% $CO_2$, 1.5% $N_2$;
- 50% Ar, 49% CO, 1% $N_2$;
- 97% $CO_2$, 3% $N_2$;
- 98% CO, 2% $N_2$;
- 70% CO, 30% $N_2$.

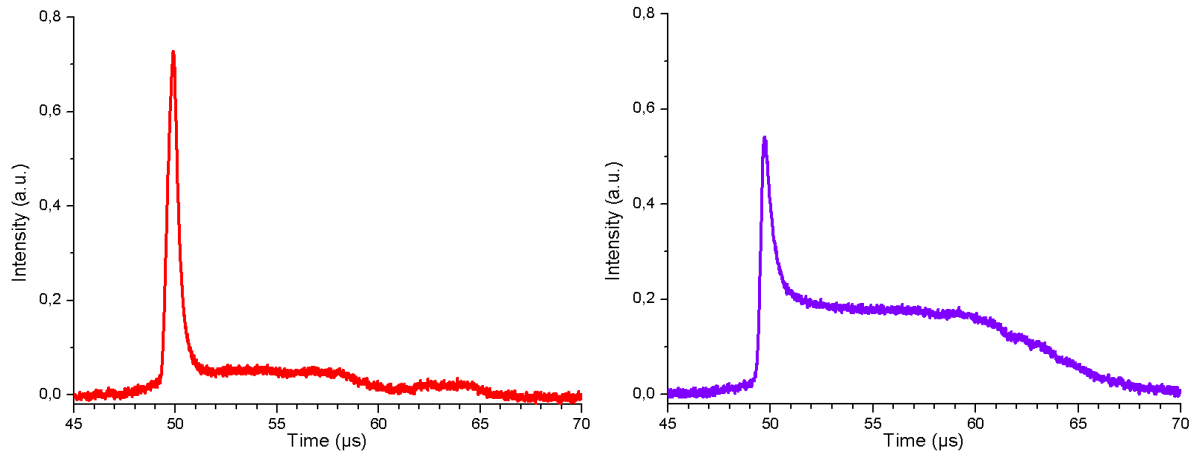

**Figure 62: Emission intensity as function of time for Case 2 of Table 6: Left CO(4+) 156-154 nm; Right: CN(violet) 381-389 nm (from [58])**

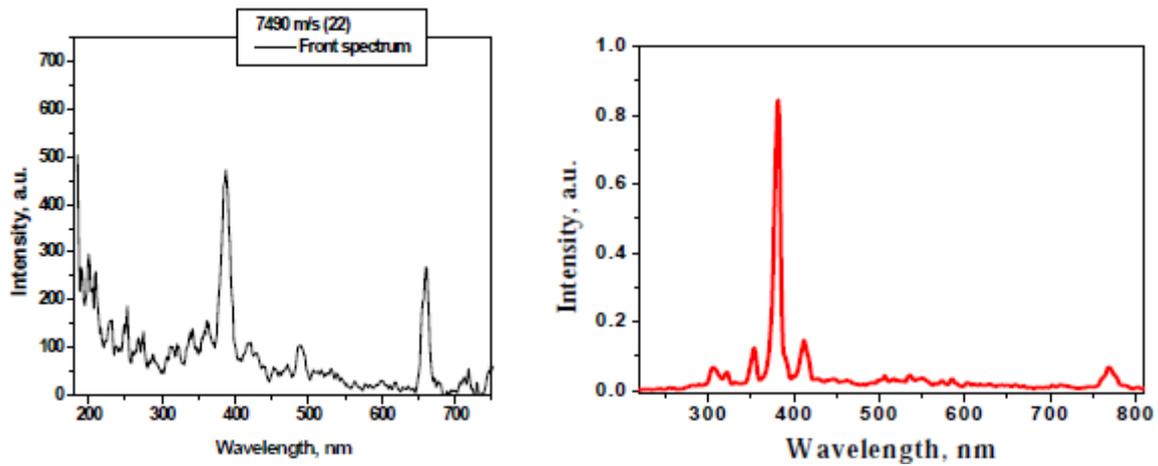

**Figure 63: Emission intensity at the shock front for Cases 3 (left) and 4 (right) of Table 6 (from [58])**

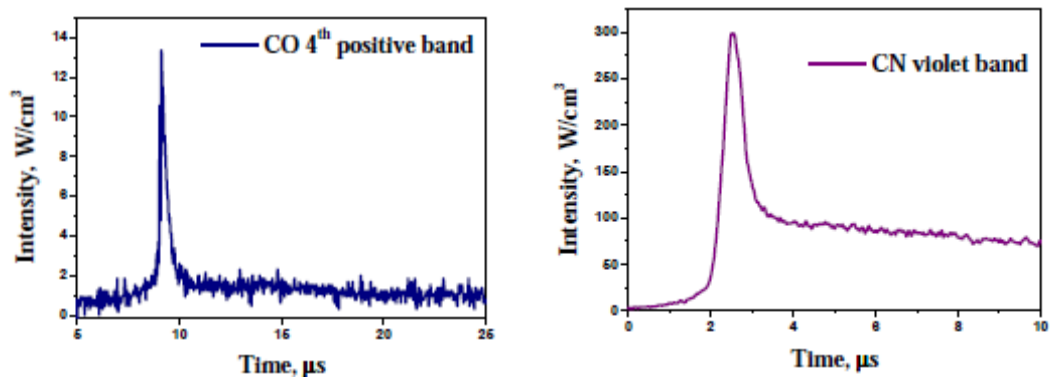

**Figure 64: CO(4+) and CN violet bands for Case 5 of Table 6 (from [58])**

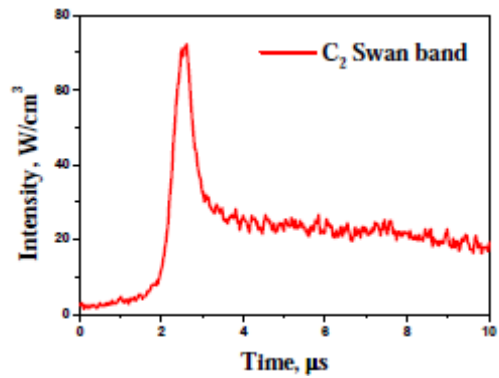

**Figure 65: $C_2$ Swan band for Case 5 of Table 6 (from [58])**

Most of these different campaigns was focused on investigating radiation into Mars atmosphere for CO(4+), CN violet, and $C_2$ Swan systems. Some of the results obtained are displayed in the following figures. A streak image obtained for the Case 1 of Table 6 is shown in Figure 61. The emission intensities obtained for Case 2 of Table 6 for CO(4+) and CN violet bands are plotted in Figure 62. Emission intensity at the shock front for Cases 3 and 4 of Table 6 as function of the wavelength is shown in Figure 63. Intensity as function of time for CO(4+), CN violet and $C_2$ Swan bands for Case 5 of Table 6 are plotted in Figure 64 and in Figure 65 .

## 5.6 Synthesis

The experimental results obtained from propulsion studies focuses on plume radiation. If some data covers the IR range, they are obtained at high-pressure levels, much higher than those encountered during the high altitude part of planetary entries when radiation can be important. However, some aspects of these studies such as measurement techniques and instrumentation are relevant for future tests to be carried out in ESTHER. As a consequence, future developments of experimental approaches in this area shall be surveyed for assessing their interest for future shock-tube campaigns.

The ground test results obtained for Venus entry conditions are resumed in Table 7 (see also Figure 7), in which the test conditions, facility, dataset available and references are reported. All datasets have been collected in EAST shock-tube. Results have been obtained for three different conditions and covers VUV to NIR wavelengths. For the time being, there is no other way, that their numerical reconstruction, to check these data sets. As a consequence, experimental results obtained in another facility for similar conditions would be of high interest for improving our knowledge on radiation during Venus entry.

| Pressure (Torr) | Velocity (km/s) | Facility | Data | Reference |
|---|---|---|---|---|
| 0.5 | 11.4 | EAST | VUV to IR | 1, 17, 18 |
| 0.5 | 10.6 | EAST | VUV to IR | 1, 17, 18 |
| 1 | 9.5 | EAST | VUV to IR | 1, 17, 18 |

**Table 7: Venus entry conditions investigated with shock-tube experiments**

A large number of experimental test campaigns has been focused on radiation measurements for Mars entry. Shock-tube facility maps for KAIST, EAST, X2, TCM2, and VUT-1 (MIPT), TC2 test conditions, and different tests carried out in these facilities are reported in Figure 17. The test conditions run in several facilities for which the most detailed results can be found in the literature are listed in Table 8. An emphasis has been put on conditions for which VUV and/or IR measurements have been performed. For some test conditions, spectra are not available, however they have been included in the list since other measurements providing some data of interest such as, temperature, electron density, and radiative heating, are available, and can relevant for some cross-checks. As far as VUV and IR data are considered, the most attractive datasets have been obtained in EAST, X2, and HVET. The duplication of some of the test conditions run in these ground tests shall have a priority for future test campaigns to be conducted in ESTHER. For most of the conditions resumed in Table 8, a Mars atmosphere has been considered, however for some tests performed in TCM2, and VUT-1, the Mars atmosphere composition was not always respected, with different proportions in $CO_2$ and $N_2$ or the addition of Argon. The tests carried out in KAIST shock-tube were done using a $CO$-$N_2$ atmosphere.

| Pressure (Torr) | Velocity (km/s) | Facility | Spectra | Reference |
|---|---|---|---|---|
| 0.2 | 6-9 | ADST | - | 42 |
| 0.1 | 8.6 | EAST | VUV to NIR | 26, 27 |
|  |  | X2 | UV-VIS | 37-38 |
| 0.1 | 8.3-9 | EAST | VUV to IR | 28 |
| 0.1 | 7-7.5-8 | EAST | VUV to IR | 26, 29 |
| 1 | 7-7.5-8 | EAST | VUV to IR | 26 |
| 0.25 | 6.5 | EAST | VUV to IR | 29 |
| 0.2-1 | 3-7.5 | EAST | IR | 32 |
| 0.075 | 7.6 | HVET | VIS-IR | 36 |
| 0.15 | 5.5 | HVET | VIS-IR | 36 |
| 1.5 | 7.3-7.5 | HVST | VUV to NIR | 36 |
| 1 | 7 | HVST | VUV to NIR | 34 |
| 0.1 | 8.5 | HVST | VUV to NIR | 34 |
| 1 | 2.5-8 | HVST | IR | 35 |
| 3.77 | 5.2 | KAIST | - | 47 |
| 7.3 | 4.83 | KAIST | - | 47 |
| 8.37 | 5.13 | KAIST | - | 47 |
| 0.3 | 6.3-7.6 | MSU | UV-VIS | 54 |
| 1 | 4.6-5.8 | MSU | UV-VIS | 54 |
| 1.47 | 5.8 | TCM2 | UV | 57-57 |
| 1.68 | 6.25 | TCM2 | UV | 55 |
| 1.8 | 6.17 | TCM2 | UV | 55 |
| 4.06 | 5.2 | TCM2 | VIS | 57-57 |
| 7.49 | 0.1 | VUT-1 | VUV-VIS | 58 |
| 7.06 | 0.6 | VUT-1 | VUV-VIS | 58 |
| 6.5 | 1.32 | VUT-1 | VUV-VIS | 58 |
| 6 | 1.42 | VUT-1 | VUV-UV | |
| 6.27 | 1.59 | VUT-1 | - | 55 |
| - | 8.5 | X2 | UV-VIS | 40 |
| 0.1 | 8.7 | X2 | UV-VIS | 37 |
| 1.13 | 4 | X2 | IR | 39 |
| 2.7 | 3.5 | X2 | IR | 39 |
| 02.7 | 2.9 | X2 | IR | 39 |

**Table 8: Mars entry conditions investigated with shock facilities (shock-tubes, expansion tubes) experiments, with $CO_2$-$N_2$ mixtures (for KAIST CO-$N_2$) related to Mars entry. In red Test Case 2 (of Radiation Working Group) conditions**

| Pressure (Pa) | Enthalpy (MJ/kg) | Facility | Data | Mixture |
|---|---|---|---|---|
| 3800 | 25 | CORIA ICP | VUV to VIS | $CO_2$ |
| 9000 | 8.5 | CORIA ICP | VUV to VIS | $CO_2$ |
| 2.9 to 5.3 | 4.16 to 35 | ICARE PHEDRA | VUV to NIR | 97 % $CO_2$ -3% $N_2$ |
| 130 | 10, 15, and 20 | IRS PWK3 | UV to VIS | $CO_2$ |
| 95000 | 19 | LAEPT ICP | UV to NIR | 97 % $CO_2$ -3% $N_2$ |

**Table 9: Test conditions investigated in some European plasma torches**

Several test campaigns have been performed recently, in Europe, for investigating radiation in plasma torch for $CO_2$-$N_2$ mixtures. Since, there is no large shock-tube facility available in Europe fitting for exploration missions [4] (until ESTHER is operational), such test campaigns are the only way for investigating radiation in plasma flows. The test conditions and available results obtained for Mars like mixture are resumed in Table 9. Datasets cover the VUV up to the NIR, and have been collected for different enthalpies and pressures. These studies have been very important for the European radiation community. Such studies have allowed maintaining European capabilities for radiation in relation to planetary entry, and have sustained the development of instrumentation and measurement techniques. The collected datasets might be of interest for future ESTHER test campaign for comparisons and checking.

Comparisons between the radiation data obtained in EAST and X2 have also been performed in [37] for 0.1 Torr and 8.5 km/s, and a Mars atmosphere composed of 96% of $CO_2$ and 4% of $N_2$. It has to be also noted that X2 results have been compared against data obtained in EAST and HVST for Mars entry conditions in [60,61]. Since crosschecks between these datasets are already available, the experimental investigation of the corresponding flow conditions shall be a priority for ESTHER.

In the perspective of comparisons, the test campaign to be conducted in ESTHER shall be focused on the following points:

- Conditions already run in the frame of ESA activities in VUT-1 and TCM2;
- Test conditions already cross checked in X2, EAST, and HVST;
- Test conditions for which VUV and NIR radiative spectra are available;
- Run Test Case 2 conditions.

## 6 Conclusions

A survey of the available results for $CO_2$ radiation data in ground facilities has been carried out. Experimental data obtained for propulsion applications have been considered, though, most of the efforts have been concentrated on the tests performed in shock-tubes and expansion tubes, for Mars and Venus entries. Results obtained for propulsion purpose did not show a strong interest for future shock-tube campaigns; however, future results from ESTHER could be of interest for propulsion. Additionally, the instrumentation and measurement techniques existing or to be developed in this field could benefit to future shock-tube development.

The emphasis has been put on VUV and NIR available datasets since dedicated instrumentation is going to be implemented in ESTHER for such wavelength ranges. A large number of datasets are available, particularly for Mars entry, this for a several facilities. A synthesis has been attempted in order to provide a first selection of the most attractive test conditions, to be run in ESTHER. This shall allow future comparisons with ESTHER results as soon as they are available.

## Acknowledgements

The research leading to these results has been partially supported by Fluid Gravity Engineering and Instituto Superior Tecnico of Lisbon under European Space Agency TRP on Characterisation of Radiation for High-speed Entry.